\documentclass[trackchanges]{aastex631}

\newcommand{\fermi}{\mbox{\textit{Fermi}-GBM~}}
\newcommand{\ferminosp}{\mbox{\textit{Fermi}-GBM}}

\newcommand{\sgrone}{SGR\,J1935+2154~}
\newcommand{\sgrtwo}{SGR\,J1550--5418~}
\newcommand{\sgronenos}{SGR\,J1935+2154}
\newcommand{\sgrtwonos}{SGR\,J1550--5418}

\usepackage{comment, soul}
\usepackage{amsmath, ulem}
\usepackage{multirow}
\usepackage{float}

\shorttitle{17-year GBM Magnetar Burst Catalog}
\shortauthors{M. Godwin}
\begin{document}

\title{17 Yr of Magnetar Bursts Observed with the \textit{Fermi} Gamma-ray Burst Monitor}

\correspondingauthor{Matt Godwin}
\email{msg0028@uah.edu}

\author[0000-0002-2926-0469]{Matt Godwin}
\affiliation{Department of Space Science, University of Alabama in Huntsville, Huntsville, AL 35899, USA}

\correspondingauthor{\"Ozge Keskin}
\email{ozgekeskin@sabanciuniv.edu}

\author[0000-0001-9711-4343]{\"Ozge Keskin}
\affiliation{Sabanc\i~University, Faculty of Engineering and Natural Sciences, \.Istanbul 34956 T\"urkiye}

\author[0009-0000-9126-7824]{Mustafa Demirer}
\affiliation{Sabanc\i~University, Faculty of Engineering and Natural Sciences, \.Istanbul 34956 T\"urkiye}

\author[0000-0000-0000-0000]{Yi-Jing Yi}
\affiliation{School of Physics and Astronomy, Beijing Normal University, Beijing 100875, China}

\author[0000-0002-0633-5325]{Lin Lin}
\affiliation{Institute for Frontiers in Astronomy and Astrophysics, Beijing Normal University, Beijing 100875, China}
\affiliation{School of Physics and Astronomy, Beijing Normal University, Beijing 100875, China}

\author[0000-0002-7150-9061]{Oliver J. Roberts}
\affiliation{Science and Technology Institute, Universities Space and Research Association, 320 Sparkman Drive, Huntsville, AL 35805, USA}
\affiliation{Physics, School of Natural Sciences, University Road, University of Galway, Galway, H91 TK33, Ireland}

\author[0000-0002-5274-6790]{Ersin G\"o\u{g}\"u\c{s}}
\affiliation{Sabanc\i~University, Faculty of Engineering and Natural Sciences, \.Istanbul 34956 T\"urkiye}

\author[0000-0002-1861-5703]{Yuki Kaneko}
\affiliation{Sabanc\i~University, Faculty of Engineering and Natural Sciences, \.Istanbul 34956 T\"urkiye}

\author[0000-0003-2105-7711]{Michael Briggs}
\affiliation{Center for Space Plasma and Aeronomic Research, University of Alabama in Huntsville, Huntsville, AL, USA}
\affiliation{Space Science Department, University of Alabama in Huntsville, Huntsville, AL, USA}

\author[0000-0003-4433-1365]{Matthew G. Baring}
\affiliation{Department of Physics and Astronomy - MS 108, Rice University,
6100 Main Street, Houston, TX 77251-1892, USA}
% \email{baring@rice.edu}

\author[0000-0002-2942-3379]{Eric Burns}
\affiliation{Department of Physics \& Astronomy, Louisiana State University, Baton Rouge, LA 70803, USA}
% \email{ericburns@lsu.edu}

\author[0000-0001-7128-0802]{David M. Palmer} 
\affil{Los Alamos National Laboratory, New Mexico Consortium}

\author[0000-0001-9556-7576]{Boyan~A.~Hristov}
\affiliation{Center for Space Plasma and Aeronomic Research, University of Alabama in Huntsville, Huntsville, AL, USA}

\author[0000-0002-2531-3703]{Bagrat Mailyan}
\affiliation{Department of Aerospace, Physics and Space Sciences, Florida Institute of Technology, Melbourne, FL 32901, USA}

\author[0000-0002-0380-0041]{Christian Malacaria}
\affiliation{INAF Osservatorio Astronomico di Roma, via di Frascati 33, I–00078, Monteporzio Catone, Roma, Italy}

\author[0000-0002-0786-7307]{Eva Palafox}
\affiliation{Instituto Nacional de Astrofísica, Óptica y Electrónica, Luis Enrique Erro $\#$1, Tonantzintla, Puebla 72840, México}

\author[0000-0003-1443-593X]{Chryssa Kouveliotou}
\affiliation{Department of Physics, George Washington University, 725 21st Street NW, Washington, DC, 20052, USA}

\author[0000-0001-9149-6707]{Alexander J. van der Horst}
\affiliation{Department of Physics, George Washington University, 725 21st Street NW, Washington, DC, 20052, USA}

\author[0000-0002-2149-9846]{Peter Veres}
\affiliation{Department of Space Science, University of Alabama in Huntsville, Huntsville, AL 35899, USA}
\affiliation{Center for Space Plasma and Aeronomic Research, University of Alabama in Huntsville, Huntsville, AL, USA}

\author[0000-0002-7991-028X]{George Younes}
\affiliation{Astrophysics Science Division, NASA Goddard Space Flight Center, Greenbelt, MD 20771, USA}
\affiliation{Center for Space Sciences and Technology, University of Maryland, Baltimore County, Baltimore, MD 21250, USA}
\email{}

%%%%%%%%%%%%%%%%%%%%%%%%%%%%%%%%%%%%%%%%%%%%
%%%%%%%%%%%%%%%%%%%%%%%%%%%%%%%%%%%%%%%%%%%%
%%%%%%%%%%%%%%%%%%%%%%%%%%%%%%%%%%%%%%%%%%%%
%%%%%%%%%%%%%%%%%%%%%%%%%%%%%%%%%%%%%%%%%%%%

\begin{abstract}

The \textit{Fermi} Gamma-ray Burst Monitor (GBM) has been in operation for over 17 years, during which it has observed more than a thousand bursts from soft gamma repeaters (SGRs), also known as magnetars. Serving as a laboratory for extreme physics, magnetars are a sub-family of neutron stars characterized by extreme magnetic field strength, observed through a combination of persistent and short transient emission across the electromagnetic spectrum. We present the comprehensive GBM catalog of SGR short bursts which supersedes the 5-year catalog of \cite{Collazzi2015}. The new catalog contains 1254 SGR short bursts observed over 17 years, providing the longest uninterrupted, high-sensitivity all-sky monitoring of magnetar bursts with unprecedented spectral and temporal resolution. Our catalog contains bursts from 17 unique Galactic sources, with major contributions by bursts from \sgrone and \sgrtwonos. We present overall characteristics of these bursts, such as the durations, spectral parameters for various photon models, fluxes, as well as their comparison with recently published catalogs of other missions and the previous GBM magnetar catalog. The machine readable catalog, as well as burst spectra and response files are made publicly available for the community.

\end{abstract}
%%%%

%%%%%%%%%%%%%%%%%%%%%%%%%%%%%%%%%%%%%%%%%%%%
%%%%%%%%%%%%%%%%%%%%%%%%%%%%%%%%%%%%%%%%%%%%
%%%%%%%%%%%%%%%%%%%%%%%%%%%%%%%%%%%%%%%%%%%%
%%%%%%%%%%%%%%%%%%%%%%%%%%%%%%%%%%%%%%%%%%%%

\section{Introduction} \label{sec:intro}

Magnetars are highly magnetized neutron stars that produce very luminous ($\sim$$10^{35}-10^{47}$ erg\,s$^{-1}$), energetic %($\gtrsim$$10^{36}-10^{46}$ erg), 
X-ray bursts, powered exclusively by the decay and evolution of their strong  magnetic fields ({$B \sim 10^{14} - 10^{15}$~G) \citep{TD92,Paczynski1992,TD95,TD96}. They are widely believed to form during the core-collapse of massive stars ($\sim$20~$M_{\odot}$), although the precise conditions required to generate their extreme magnetic fields remain uncertain. They were discovered initially in 1979, when the KONUS instruments on the Venera 11 and 12 satellites detected a series of short, energetic events from a single source~\citep{Mazets1979b,Mazets1979a}. Originally classified as short gamma-ray bursts (GRBs), \citet{Mazets1981} subsequently argued that their phenomenology was not aligned with conventional GRBs in both time profiles and spectral characteristics. These events are now thought to originate from soft gamma repeaters (SGRs), slowly rotating, isolated X-ray pulsars embedded in supernova remnants suggesting that they are young neutron stars~\citep{Kouv98}. Shortly after the \citet{Mazets1979b} report, the Einstein Observatory independently observed an X-ray pulsar (1E 2259+586) in a galactic supernova remnant (G109.1$-$1.0), which was later found to have a spin period of 6.98 s ~\citep{Fahlman1981}.
These sources (later labeled Anomalous X-ray Pulsars or AXPs), exceeded the luminosities expected from rotational energy, and showed no evidence of binary companions \citep{Mereghetti1995}, indicating they were powered by some other phenomenon like a strong magnetic field~\citep{Vasisht1997}. 
Subsequent measurements of the large spin-down rates for both SGRs and AXPs indicated that both source classes are likely powered by the evolution of their extremely strong magnetic fields, while the discovery of SGR-like bursts from the AXP 1E~2259+586 provided compelling evidence that the two source classes are manifestations of the same underlying phenomenon, unifying them under the magnetar model~\citep{Kaspi2003,Kouv98,Kouveliotou1999,Woods2006,Mereghetti2008}.
Recent population studies suggest that magnetars represent a small but significant subset of the neutron star population \citep{Keane2008,Beniamini2019,Araujo2026}. Observational properties of the currently known magnetar population are summarized in the McGill Magnetar Catalog \citep{Olausen2014}, while the broader neutron star population is cataloged in the ATNF Pulsar Catalogue \citep{atnf-catalog}.

Magnetars emit persistent radiation detected predominantly in the X-ray and high-energy bands \citep{Gotz2006,Kuiper2008,Enoto2010}.
Recent observations with the Imaging X-ray Polarimetry Explorer (IXPE) revealed that their X-ray emission is highly polarized, independently confirmed the presence of ultra-strong magnetic fields, and provided insights into the emission mechanism \citep{Turolla24}. Magnetars also display a diverse range of transient, high energy phenomena during heightened activity known as ``outbursts."\footnote{Magnetar outbursts refer to abrupt increase in persistent X-ray emission usually in connection with bursts and gradual decline back to the quiescent level within months to years, see e.g., \citealt{Cotizelati18}.} Short duration bursts are the most common ($\sim$10$^{-2}$ s to $\sim$1 s;~\citealt{Gogus01}), typically releasing total energies ranging from $10^{38} - 10^{41}$~erg. Occasionally, magnetars release longer (1$-$50~s), more energetic ($10^{41} - 10^{43}$~erg) bursts called intermediate events \citep{Mereg09}. Very rarely, magnetars emit giant flares with total energies $\gtrsim$10$^{44}$ erg \citep{Hurley1999,Palmer2005}. During these periods of heightened activity, they may also exhibit abrupt variations in their spin behavior, known as glitches \citep{Kaspi2003,Hu2024Natur} that may be accompanied by other transient phenomena such as Fast Radio Bursts~\citep{chime20,Bochenek2020,mereghetti2020,Li2021,Ridnaia2021,Hu2024}. The theoretical magnetar model~\citep{TD92,TD95,TD96} attributes the persistent and transient high energy emission of these sources to magnetic energy stored within the neutron star and its magnetosphere. Magnetic stresses acting on the crust can produce fractures and reconnection events, triggering burst activity within the X-ray and gamma-ray regime.

Continued monitoring of magnetars across the electromagnetic spectrum has been carried out by numerous space-based observatories including Neil Gehrels \textit{Swift} Observatory (\textit{Swift}; \citealt{gehrels2004}), INTEGRAL (International Gamma-Ray Astrophysics Laboratory; \citealt{Winkler2003}), \textit{NICER} \citep{nicer2016}, and the \textit{Fermi} Gamma-ray Space Telescope \citep{Meegan2009,Atwood_2009}. The results from these surveys have been published in catalogs: The \textit{NICER} magnetar bursts catalog~\citep{Chu2026} contains over 1100 bursts, mostly dominated by \sgronenos. From their extensive spectral analysis in the limited energy range of 0.5–8 keV, \citet{Chu2026} states that bursts with higher fluences tend to have harder spectra, and correlations between burst duration and spectral parameters are weak. Another catalog of magnetar bursts observed by the Imager on Board the INTEGRAL Satellite (IBIS) data contains over 1300 events, spanning two decades of data taking in the 15$-$1000 keV range \citep{Pacholski2026}. 
The IBIS catalog is dominated by events from \sgrtwo (also known as 1E\,1547.0$-$5408), SGR 1806$-$20, and \sgronenos. \citet{Pacholski2026} presents extensive burst characteristics (duration and spectral shapes), as well as correlative studies, such as an anti-correlation between $E_{\rm peak}$ and fluence for SGR 1806$-$20. 

The first comprehensive \fermi catalog of magnetar bursts during the first five years of operation (2008$-$2013), was presented in \citet{Collazzi2015}. The catalog provided durations, fluences, peak fluxes, and spectral parameters for over 400 bursts, primarily from active sources (SGRs J1550--5418, 1806--20, and 0501+4516), along with 19 events from unconfirmed sources. Since 2013, \fermi has continued its uninterrupted operation and has significantly expanded the database of magnetar bursts.

The catalog presented here contains a total of 1254 magnetar short bursts observed with \fermi between July 2008 and December 2025. It includes SGRs that were autonomously detected (``triggered") by Fermi GBM, as well as SGR bursts identified in the data readout up to 300\,s after the trigger time of an SGR burst. Each of the triggered events has its association with a magnetar confirmed by a combination of its localization, whether the event was detected by another instrument, temporal association with a source in known outburst, manual light-curve inspection, and energetics. All triggered events contained within the earlier catalog \citep{Collazzi2015} are also listed here with new spectral analysis, except for those that were re-examined and deemed unrelated to magnetar activity.

The paper is structured as follows: The next section (\S\ref{sec:obs}) describes the data used in this study and the methodologies employed in searching for untriggered bursts, localization of events, and burst characterization, namely, the duration and the spectral analysis. In \autoref{sec:res}, we present our results and discuss them in \autoref{sec:disc}.

%%%%%%%%%
\section{Observations and Methodology}\label{sec:obs}

\subsection{Data Selection}
The Gamma-ray Burst Monitor (GBM) is an all-sky monitoring instrument onboard the \textit{Fermi} Gamma-ray Space Telescope, the secondary instrument to the Large Area Telescope (LAT). It is equipped with 12 Thallium-doped Sodium Iodide (NaI) detectors (numbered 0$-$9,a,b), which are sensitive to photons in the 8$-$900~keV energy range, and two Bismuth Germanate (BGO) detectors that are sensitive to photons between 280 keV and 40 MeV. The NaI detectors are arranged into four clusters of three detectors, placed at the corners of the spacecraft, providing reasonably uniform sky coverage that is not Earth occulted~\citep{Meegan2009}.

GBM provides several data types in continuous mode and trigger mode. Triggered data types provide higher resolution data and include CTIME, CSPEC, and triggered Time-Tagged Event (TTE) data. The TTE data records individual photon information for 30~s or $\sim$500k events pre-trigger, and 300\,s post-trigger time at a time resolution of $\sim$ 2 $\mu$s, over 128 energy channels, with a maximum rate in all detectors of 375 kHz.
Since an update of the flight software in November 2012, a Continuous Time Tagged Event (CTTE) data type has become available with the same time and energy resolutions as the trigger TTE data type. However, we only use the triggered TTE data in this study because they provide more reliable association between the observed bursts and their source and are available for the full time span of the catalog.

Since the energy range of the BGO detectors exceeds the expected soft energy range for SGR bursts ($<$300~keV), we only use the NaI detectors for our analyses. Due to Earth occultation, field-of-view obstruction by LAT, and the location of these events, only a subset of these detectors can observe a specific event at any given time. For each burst, we used only detectors with an angle between the source direction and the detector normal of less than 60\textdegree, to ensure a large effective area and a well-modeled response for these events. We also excluded any detectors occulted by the Earth or blocked by LAT. 
%\citet{Collazzi2015} \hlok{notes that in previous studies of GBM bursts, this viewing angle criterion varies from 40 to 60\textdegree, but this does not have a significant effect on the temporal and spectral results.  $-->$ Do we need this sentence?}
We included all triggered events identified as SGRs by the \fermi team (688 events) and the untriggered events identified within the associated triggered TTE data. Since this is a catalog of SGR short bursts, we excluded a peculiar episode consisting of longer events previously reported as a ``burst forest" \citep{Kaneko2021} from the most prolific magnetar, \sgrone (occurring 380$-$390 s and 400$-$420 s after trigger bn200427768).

\subsection{Untriggered Event Search}
\label{subsec:untriggered}

\fermi relies on onboard trigger algorithms \citep{vonKienlin2012}
and subsequent manual verification on the ground to effectively detect and identify magnetar bursts \citep{kaneko2026}. The first data type produced during an onboard trigger is the TRIGDAT, trigger data type, designed to contain the minimum amount of data required for rapid on-ground characterization and localization of triggers. It has 8-channel, pre-binned lightcurve data for each of the 14 detectors along with spacecraft position and attitude information for each bin. During the writing and readout of TRIGDAT data, which lasts around 600 seconds, GBM is unable to re-trigger on other events. The higher resolution CTIME, CSPEC and TTE data are produced hours later for analysis. Magnetar short bursts are generally repetitive and clustered, particularly during active episodes when the burst rate is high, which results in some bursts being recorded during the 600 s TRIGDAT readout time that fail to trigger GBM. To search for and identify these untriggered bursts, we applied the Bayesian block method \citep{scargle2013} in the triggered event data as follows. We first extracted events in the 10$-$100 keV range from each NaI detector's triggered TTE data to generate light curves with a time resolution of 8 ms, and then reconstructed the light curves via Bayesian blocks~\citep{lin2020b}.
Blocks with durations shorter than 1 s are defined as burst blocks, while temporally adjacent blocks longer than 1 s are taken as background blocks. The mean count rate of the background blocks serves as the background rate. We further applied the criteria of count rates exceeding the background rate to identify burst blocks, and merge consecutive burst blocks to form candidate bursts.

\subsection{Event Localization}

The events in this catalog were initially localized by a Burst Advocate (Human-in-the-Loop; HitL) using the GBM TRIGDAT, trigger data, by GBM's internal flight software, or by another satellite with better localization capability (i.e., arcminute resolution using the Burst Alert Telescope (BAT) on \textit{Swift}; \citealt{Barthelmy2005}). Many of the localizations that lacked confirmation from Swift BAT were manually reanalyzed (HitL). For the majority of events in this catalog, they were only localized by the GBM instrument and HitL team. GBM performs localizations using the difference in the count rates among the 12 NaI detectors to map the localization contour on the unocculted sky. For GRBs, these are done using one spectral model over 50$-$300 keV by the Flight Software and on the ground by RoboBA~\citep{Goldstein2020} and the HitL~\citep{Avk2020}. The location uncertainties from those calculations are statistical errors, which have additional systematic error from the HitL locations that are modeled as a core-plus-tail model, extending from 3.7\textdegree (for 90\% of bursts) to 10\textdegree~\citep{Connaughton2015}. An improvement to the localization algorithm, RoboBA, improved the systematic uncertainty to 1.8\textdegree~for 52\% of GRBs, and 4.1\textdegree~for the remaining 48\% GRBs~\citep{Goldstein2020}. 
For SGR bursts, localizations are determined by a HitL using data predominantly from the soft channels (8--50 keV) and a Band spectral model that is folded into the data with $\alpha$, $\beta$, $E_{\rm peak} = -1.9, -3.7, 70$ keV, respectively. These localizations are shown in Table \ref{tab:main_table}. We note while the reported burst locations correspond to the localization solution with the lowest chi-square -- the statistically best-fitting solution, these locations do not necessarily have the smallest uncertainty. In some cases, simultaneous observations by instruments with superior localization capabilities (e.g.\,\textit{Swift}-BAT) show that the derived positions can be significantly offset from the true source location. This most commonly occurs for bursts near the Earth limb, where the localization is strongly affected by background Earth-limb albedo. Bursts occurring during periods of enhanced soft emission (i.e., particle or solar activity), may likewise have somewhat biased locations and underestimated or misrepresented uncertainties. Despite this, the localizations performed here are well constrained for the majority of bursts, which is shown in Figure \ref{fig:1935locmap} for \sgrone bursts. This figure also shows the distribution of the localization offsets to the accurate source location. The full-width at half maximum (FWHM) of RA measurements for triggered bursts is 8.3$^\circ$, and that of untriggered events is 12.9$^\circ$. Similarly, the FWHM of Dec measurements are 8.9$^\circ$ and 15.9$^\circ$ for triggered and untriggered bursts, respectively. As expected, untriggered events are generally weaker than triggered ones, resulting in a larger dispersion in their localizations. 

\begin{figure}[htbp!]
    \centering
    \includegraphics[width=0.8\linewidth]{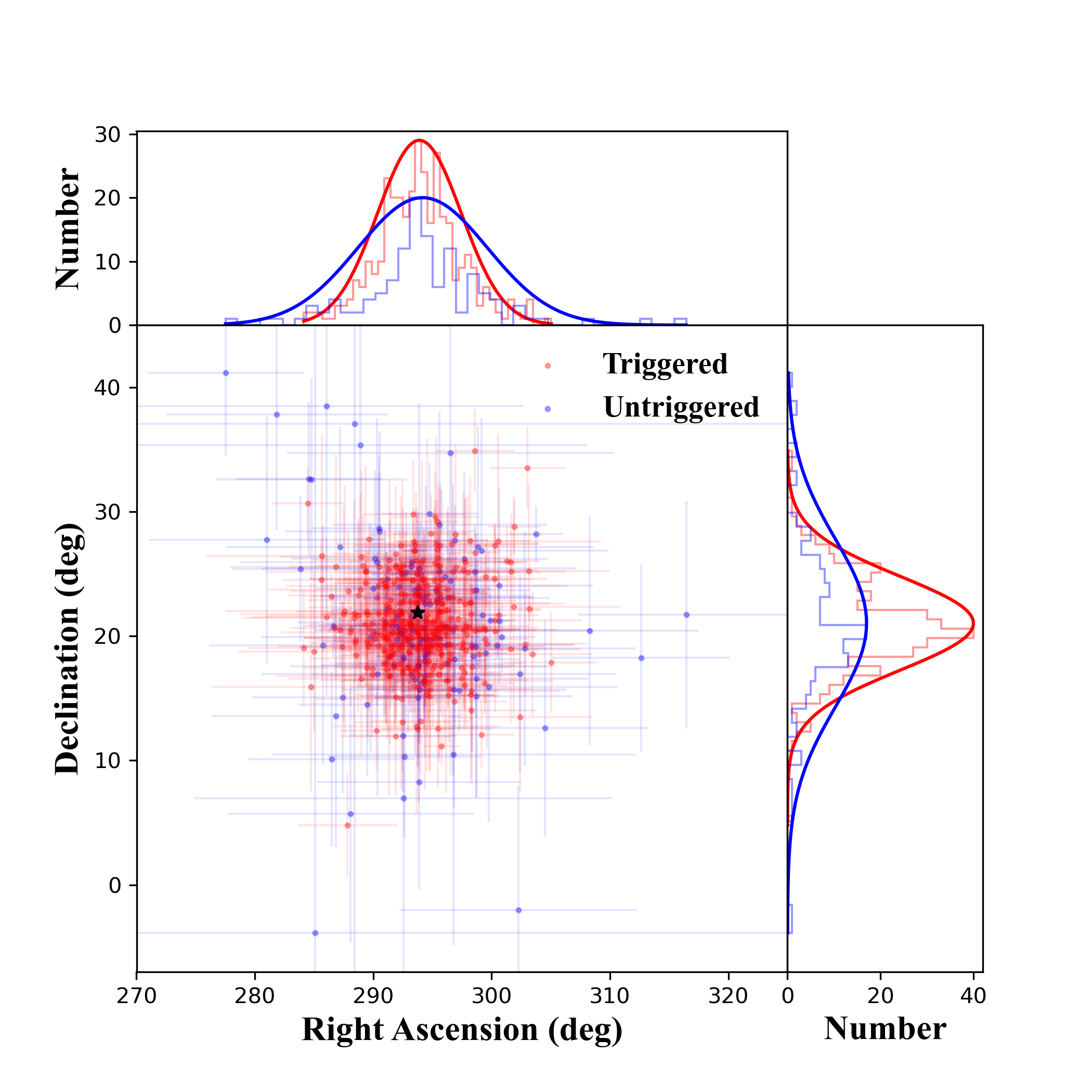}\\
    
    \caption{\textbf{[Main]} Localization of triggered (red) and untriggered (blue) \sgrone bursts with their 1$\sigma$ uncertainties.  The star denotes the accurate location of the magnetar. \textbf{[Top]} The distributions of Right Ascension for triggered (red) and untriggered (blue) events. \textbf{[Right]} The distributions of Declination for triggered (red) and untriggered (blue) events.}
    \label{fig:1935locmap}
\end{figure}

Although every effort has been made to obtain the most accurate GBM localization possible, the localization of very short and soft bursts is limited by the 64~ms time resolution and lower energy range of the trigger data. For events with only one or two significant bins above background within the 8-200~keV, localization alone is often insufficient to uniquely identify the source. In these cases, we use contemporaneous outburst activity of nearby magnetars (i.e., Figure ~\ref{fig:bursts_barplot}) as additional evidence to support the association of the burst with a known magnetar source.

\subsection{Event Duration}

To measure burst durations more precisely, we incorporated localization information. We combined photons in the 8–200 keV range from NaI detectors with incident angles to the source of less than 60 degree and without any blockage by other parts of the satellite, and generated light curves with a time resolution of 1 ms. These light curves are then processed using the Bayesian block method \citep{scargle2013}. 
Following the same criteria as those used for untriggered event search, we identify burst blocks and background blocks, and merge consecutive burst blocks to form a complete burst profile. The duration of each burst ($T_{bb}$) is defined as the interval from the start time of the first burst block to the end time of the last burst block for an event of interest. 

We formed logarithmic distributions of $T_{bb}$ durations for all 1254 bursts in our sample, as well as for the individual prolific sources; \sgronenos, \sgrtwonos, SGR 1806$-$20 and 1E 1841$-$045, along with bursts from other 13 sources (see Figure \ref{fig:duration_dist}). We then fit each of these logarithmic distributions with Gaussian functions, which yield parameters with acceptable statistics as presented in Table \ref{tab:burst_dur_fits}. The mean values of $T_{bb}$ durations of \sgrone and \sgrtwo are remarkably similar, around 200 ms, while that of SGR 1806$-$20, 1E 1841$-$045 and all others are significantly lower and  in agreement with each other. 
%For example, the mean $T_{bb}$ of SGR 1806$-$20 is about 4.7$\sigma$ apart from that of \sgrtwonos, and the mean of 1E 1841$-$045 durations is about 6.6$\sigma$ apart. On the other end, the mean durations of 1806$-$20, 1E 1841$-$045 and all others are in agreement with each other.

\begin{figure}[htbp!]
    \centering
    \includegraphics[width=0.6 \textwidth, trim=85 92 70 265, clip] {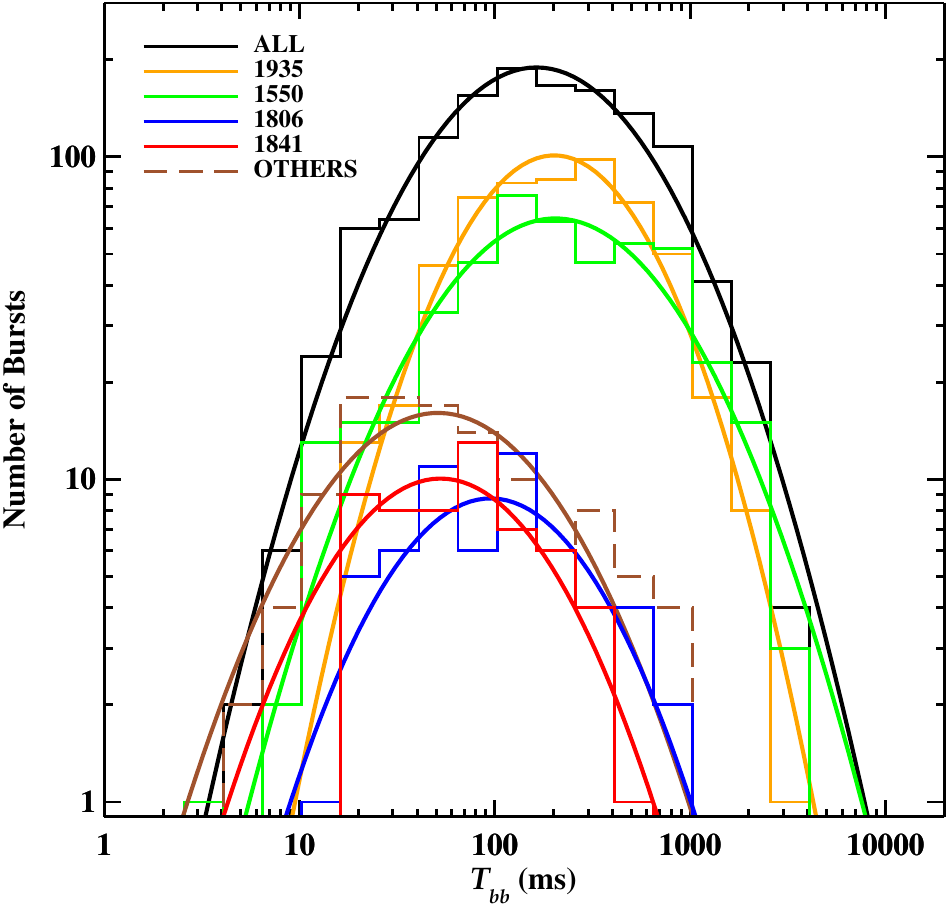}
    \caption{$T_{bb}$ distribution of all sample (black), SGR J1935+2154 (orange), SGR J1550$-$5418 (green), SGR 1806$-$20 (blue), 1E 1841$-$045 (red), and others (brown, long-dashed). The best-fitting Gaussian model is overlaid on each distribution. The logarithmic y-axis is used solely to better illustrate the differences in the duration distributions among the magnetars.}
    \label{fig:duration_dist}
\end{figure}

\begin{deluxetable*}{lccccc}
\tablecaption{Gaussian fit parameters for the logarithmic distributions of $T_{bb}$.}
\label{tab:burst_dur_fits}
\tablewidth{0pt}
\tablehead{
\colhead{\bf Source} & \colhead{\bf Number of Bursts} & \colhead{\bf $\mu$} & \colhead{\bf $\sigma$} & \colhead{\bf Mean (ms)} & \colhead{\bf $\chi^2$/dof}}
\startdata
ALL             & 1254 & $2.21 \pm 0.02$ & $0.52 \pm 0.01$ & 162.2   & 22.2/13 \\
SGR J1935+2154  & 567  & $2.30 \pm 0.02$ & $0.44 \pm 0.01$ & 199.5   & 15.2/10 \\
SGR J1550$-$5418 & 459  & $2.31 \pm 0.03$ & $0.54 \pm 0.02$ & 204.2  & 21.3/12 \\
SGR 1806$-$20   & 57   & $1.98 \pm 0.08$ & $0.49 \pm 0.07$ & 95.5   & 5.3/7 \\
1E 1841$-$045   & 56   & $1.72 \pm 0.14$ & $0.51 \pm 0.11$ & 52.5  & 2.7/5 \\
OTHERS          & 115  & $1.71 \pm 0.06$ & $0.54 \pm 0.06$ & 51.3   & 8.5/9
\enddata
\tablecomments{\\
The parameters $\mu$ and $\sigma$ denote the centroid and standard deviation of the Gaussian fits to the $\log_{10}(T_{\rm bb}/{\rm ms})$ distributions, respectively. Errors are indicated in 1$\sigma$ confidence level.
}
\end{deluxetable*}

%\begin{deluxetable*}{lcccc}
%\tablecaption{Gaussian fit parameters for $T_{bb}$ distributions}
%\label{tab:burst_dur_fitss}
%\tablewidth{0pt}
%\tablehead{
%\colhead{\bf Source} & \colhead{\bf Number of Bursts} & \colhead{\bf Mean Duration (ms)} & \colhead{\bf  Sigma (ms)} & \colhead{\bf $\chi^2$/dof}}
%\startdata
%ALL & 1254 & $163.6 ^{+5.8}_{-5.6}$ & $3.3 \pm 0.1$ & 22.2/13 \\
%SGR J1935+2154 & 567 & $201.3 ^{+8.9}_{-8.6}$ & $2.7 \pm 0.1$ & 15.2/10 \\
%SGR J1550$-$5418 & 459 & $204.4 ^{+13.2}_{-12.4}$ & $3.5 \pm 0.2$ & 21.3/12 \\
%SGR 1806$-$20 & 57 & $95.2 ^{+18.7}_{-15.6}$ & $3.1 \pm 0.5$ & 5.3/7 \\
%1E 1841$-$045 & 56 & $52.7 ^{+19.3}_{-14.2}$ & $3.2 ^{+1.0}_{-0.7}$ & 2.7/5 \\
%OTHERS    & 115 & $51.1 ^{+7.4}_{-6.4}$   & $3.5 ^{+0.5}_{-0.4}$ & %8.5/9 \\
%\hline
%\enddata
%\tablecomments{\\
%Errors are indicated in 1$\sigma$ confidence level.
%}
%\end{deluxetable*}

\subsection{Spectral Analysis}\label{sec:spec_analysis}

Time-integrated spectral analysis was performed for each burst in the catalog using Xspec \citep[version 12.14.1;][]{Arnaud1996} with Castor statistics \citep[$C$-stat;][]{Castor}. For the analysis, TTE data were used from typically the three brightest NaI detectors with detector-to-source angles less than $60^{\rm o}$ at the time of the event. We excluded the detectors that were partially or fully blocked by other parts of the spacecraft, as identified using the GBMBLOCK software provided by the \fermi team. While extracting time-integrated spectra, we employed data accumulated during the entire $T_{bb}$ duration presented in the previous subsection. In cases of data saturation, however, only the non-saturated time intervals were included in the spectral analysis. The Bayesian Block duration and detectors used in spectral analysis are listed in Table \ref{tab:main_table} for each burst. For the background estimation, $\sim$50 s burst-free data segments were selected before and after the burst. Detector response matrices were also generated using the GBM Response Generator\footnote{\url{https://fermi.gsfc.nasa.gov/ssc/data/analysis/rmfit/DOCUMENTATION.html}}, released by the \fermi team, for each burst based on its start time and the source location.

We fit each spectrum over the 8$-$200 keV energy range with four different photon models: a power law (PL\footnote{$f_{_{\rm PL}}(E) = A E^\gamma$}), an exponentially cutoff power law model (COMPT\footnote{$f_{_{\rm CO}}(E) = A \exp{[-E(2+\alpha)/E_{\rm peak}]}(E/50\,{\rm keV})^{\alpha}$}), single black body function (BB), and the sum of two black body functions (BB+BB).
 The four photon models were selected for the following reasons: COMPT and BB+BB are commonly used to describe magnetar short bursts while some magnetar bursts are described well with BB \citep{vonKienlin2012}.  On the other hand, PL has been used to fit magnetar bursts in combination with a blackbody \citep{Kirmizibayrak2017} or when the energy range is limited \citep{Chu2026}.  Theoretically, a single PL is not expected to well describe magnetar short bursts; however, in our case, it can still provide meaningful information about the spectra for weak untriggered bursts with low photon counts or very soft or hard events with a cutoff energy close to the energy edges ($\lesssim 20$\,keV or $\gtrsim 150$\,keV).
 
 The goodness of fit for each spectral model was evaluated using the $C$-statistic following \citet{Kaastra2017}.  For each fit, the expected $C$-stat ($C_{\rm e}$) and its variance ($C_{\rm v}$) were computed from the model predictions, and the fit was considered statistically acceptable if its $C$-stat value was within 3$\sigma$ (where $\sigma = (C\text{-stat} - C_{\rm e}) / \sqrt{C_{\rm v}}$) of the expected $C$-stat distribution. In addition, model parameters were required to be well constrained, such that the parameter value and its 1$\sigma$ uncertainty corresponded to physically viable values, for a fit to be considered acceptable. 
 
 Note that the 30$-$40 keV energy range was excluded from the analysis to avoid contamination from the iodine K-edge, which can affect the spectral statistics for bright events\footnote{\url{https://fermi.gsfc.nasa.gov/ssc/data/analysis/GBM_caveats.html}}. It should also be noted that the light curves of 35 bright events included time intervals during which the TTE data were saturated due to very high count rates (i.e., the total count rates of all 14 detectors $>$ 375 kHz; \citealt{Meegan2009}) resulting in data loss.  These events are flagged in \autoref{tab:spectral}, and the saturated time intervals were excluded from the spectral analysis. Following the fits, we computed the energy flux of each burst over the 8$-$200 keV band using the fit parameters of each photon model.

%%%%%%%%%
\section{Results}\label{sec:res}

\subsection{The Catalog}

Besides the 688 events that triggered \ferminosp, our extensive search for more magnetar bursts in the readout of GBM-trigger-data resulted in an additional 566 magnetar events. Therefore, the catalog presented here contains a total of 1254 bursts from 17 Galactic magnetars. We present the list of these bursts in \autoref{tab:main_table}, which includes not only the burst specific details, namely detection time, duration and originating source, and also the list of GBM NaI detectors whose normal angle to the corresponding source at the time of the event is less than 60$^{\circ}$ and not blocked by any components of the \textit{Fermi} satellite. 
There are generally three brightest detectors listed in Table \ref{tab:main_table}. Note that in the first column of Table \ref{tab:main_table}, MET stands for \textit{Fermi} Mission Elapsed Time. This is the number of seconds which have passed since 00:00:00 UTC on January 1, 2001. Also, certain untriggered events within the catalog have a reported localization error of 0.00. These events had the correct detector angles and temporal coincidence to confirm their origin without the need for a localization calculation. Therefore, they are simply assigned the exact, known location of the source.
The magnetar catalog is also publicly available at \url{https://magnetars.sabanciuniv.edu/sgr_catalog/}, the Fermi Science Support Center (FSSC) \url{https://fermi.gsfc.nasa.gov/ssc/data/access/gbm/sgr}, as well as in a Zenodo repository:\dataset[doi:10.5281/zenodo.21351728] {https://doi.org/10.5281/zenodo.21351728} \citep{zenodo-catalog}, which includes the lightcurves, count spectra and corresponding response files of all bursts.
%and the \textit{Fermi} Science Support Center\footnote{\url{https://fermi.gsfc.nasa.gov/ssc/data/} (will be available)}

\begin{deluxetable*}{lcccccccc}
\setlength{\tabcolsep}{3pt}
\tablecaption{List of 1254 magnetar bursts observed with \ferminosp.  The full list is available in the electronic version of the paper. \label{tab:main_table}}
\tablewidth{0pt}
\tabletypesize{\footnotesize} 
\tablehead{
    \colhead{Burst Start Time$^a$} & 
    \colhead{Burst Start Time$^b$} & 
    \colhead{Burst Name$^c$} &
    \colhead{Duration ($T_{bb}$)$^d$} & 
    \colhead{NaI Detectors$^e$} &
    \colhead{RA} &
    \colhead{Dec} &
    \colhead{Error$^f$} &
    \colhead{Source} \\
    \colhead{(\textit{Fermi} MET)} &
    \colhead{(YYMMDD UTC)} & 
    \nocolhead{} & 
    \colhead{(ms)} &
    \colhead{} & 
    \colhead{($^{\circ}$)} & 
    \colhead{($^{\circ}$)} & 
    \colhead{($^{\circ}$)} & 
    \colhead{}
}
\startdata
241101720.434$^{SW}$ & 080822 12:41:59.434 & 080822529 & 108 & 8;7;6 & 78.6  & 45.0  &  4.9 & SGR 0501+4516\\
241140778.728 & 080822 23:32:57.728 & 080822981 & 20 & 2 & 95.9  & 40.7  &  6.4& SGR 0501+4516\\
241144093.533 & 080823 00:28:12.533 & 080823020 &  77 & 4;3  & 74.3  & 50.9  &  1.7& SGR 0501+4516\\
241150299.965$^{SW}$ & 080823 02:11:38.965& 080823091 &    771 & a;b  & 70.9  & 34.8  &  2.7& SGR 0501+4516\\
241157422.756 & 080823 04:10:21.756& 080823174 &    251 & 0;1;9 & 76.3  & 35.1  &  5.1& SGR 0501+4516\\
241163795.052 & 080823 05:56:34.052 & 080823248 &    333 & 2;a  & 81.1   & 42.5  &  3.8& SGR 0501+4516\\
241167673.709$^{SW}$ & 080823 07:01:12.709 & 080823293 &    178 & 3;5;0 & 82.3   & 48.0   &  2.5 & SGR 0501+4516\\
241167866.188 & 080823 07:04:25.188 &          &     57 & 3;1;0 & 75.3  & 45.3  &  0.0 & SGR 0501+4516\\
241169975.910 & 080823 07:39:34.910 & 080823319 &    123 & 9;a;6 & 76.3   & 41.5   &  1.9 & SGR 0501+4516\\
241170949.287 & 080823 07:55:48.287 & 080823330 &    320 & 4;3;8 & 77.4  & 47.3  &  1.3& SGR 0501+4516\\
241173005.198$^{SW}$ & 080823 08:30:04.198 & 080823354 &     74 & 8;b  & 89.1   & 46.4   &  7.8 & SGR 0501+4516\\
241179497.452 & 080823 10:18:16.452 & 080823429 &    126 & 0;1;5 & 80.7  & 48.8  &  5.4& SGR 0501+4516\\
241183655.975$^{SW}$ & 080823 11:27:34.975& 080823478 &    507 & 8;4;b & 82.2  & 39.0     &  1.0 & SGR 0501+4516\\
241196187.243$^{SW}$ & 080823 14:56:26.243& 080823623 &    233 & a;b  & 67.6  & 39.8  &  8.3& SGR 0501+4516\\
241204132.565 & 080823 17:08:51.565 & 080823714 &    453 & 9;a;6 & 75.5  & 46.2  &  1.7 & SGR 0501+4516\\
241215573.997 & 080823 20:19:32.997 & 080823847 &    359 & 9;a;6 & 74.9  & 47.3   &  1.1& SGR 0501+4516\\
241227568.011$^g$ & 080823 23:39:27.011 & 080823986 &     39 & 9;7;6 & 91.4  & 46.4  &  8.5& SGR 0501+4516\\
...
\enddata
\tablecomments{ \\
    $^a$ Burst start time in \textit{Fermi} Mission Elapsed Time (MET) \\
    $^b$ Burst start time in UTC \\
    $^c$ Burst name from the \fermi Trigger Catalog. If this space is blank, the event was untriggered. \\ 
    $^d$ $T_{bb}$ refers to Bayesian Block duration. \\
    $^e$ Brightest detectors used in spectral analysis and with detector-to-source angle less than 60$^{\circ}$ and not partially or fully blocked by the spacecraft itself (ordered by brightness)\\
    $^f$ 0\textdegree  error means that an untriggered event was assigned the known location of its confirmed source. \\
    $^g$ Burst lies within a 64-ms long time bin of the triggered data. \\
    $^{SW}$ Event was also detected by Swift BAT. \\
    $^*$ Saturated bursts
}
\end{deluxetable*}

\begin{figure}[htbp!]
    \centering
    
  \includegraphics[angle=90,width=0.75\paperwidth,height=0.8\paperheight]{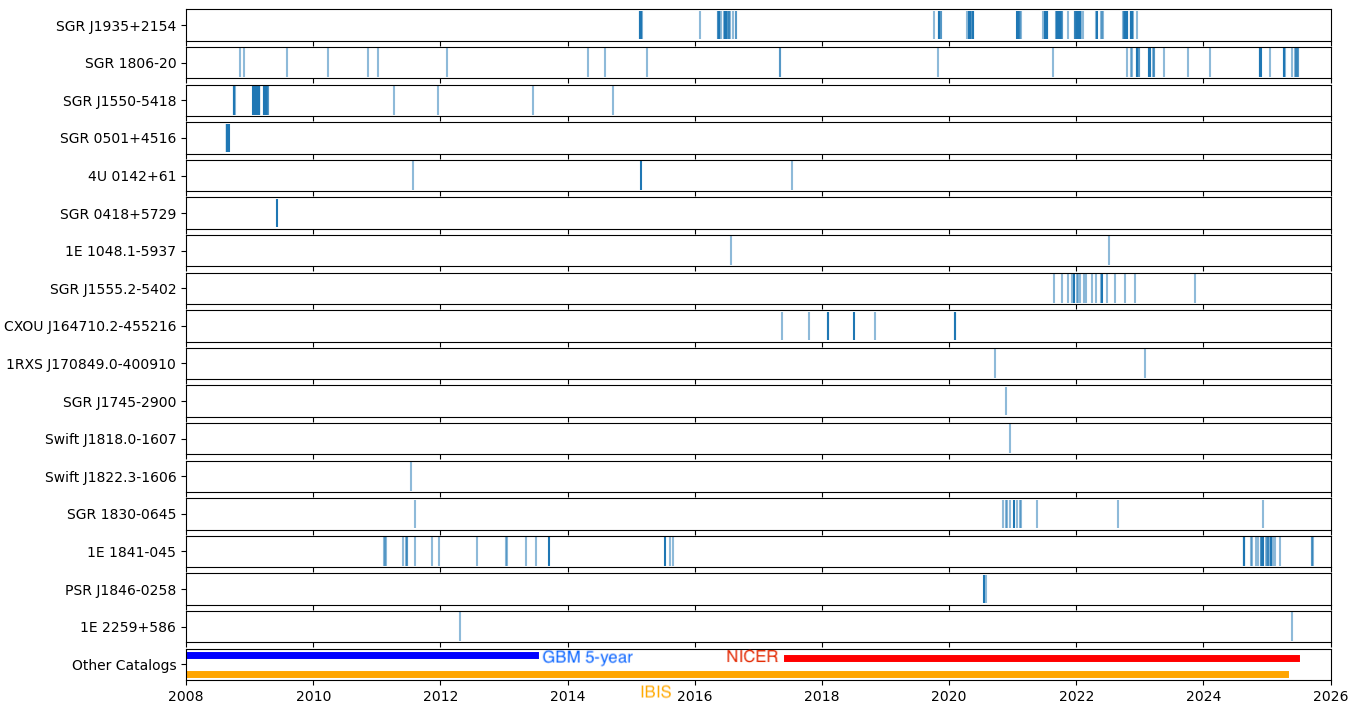}
    \caption{Occurrence times of events across the 17 year observation period. The coverage of the first GBM magnetar catalog, along with the NICER and IBIS catalogs are shown in blue, red, and orange, respectively.}
    \label{fig:bursts_barplot}
\end{figure}

We also present in Figure \ref{fig:bursts_barplot} the timeline of bursts detected from each of the 17 magnetars throughout 17 years of observing. SGR 0501+4516 became burst active about a month after GBM was commissioned in 2008 July. All of the 27 bursts from  SGR 0501+4516 were detected in a time span of 27 days starting from 22 August 2008. Following \citet{Gogus2014}, we calculated the D$_{90}$ duration of burst active episode. In particular, the onset of a burst-active episode is marked by the detection of at least five bursts from the same source within 24 hours. The episode is considered to end with the last burst preceding a quiescent interval of at least one month during which no additional bursts are detected. The D$_{90}$ duration of burst active episode is the time interval starting from the well defined onset up until 90$\%$ of bursts are detected. We find the D$_{90}$ duration of 2008 burst active episode of SGR 0501+4516 as 4.72 days. Following its activation in 2008, the source has not seen to emit any bursts neither with \fermi nor any other satellites. In Table \ref{tab:burst_active_dur}, we list the bursting episodes falling into \citet{Gogus2014} burst active epoch definition, as well as their D$_{90}$ durations. It is important to note that 1E 1841-045 is the only persistent magnetar within this list, while the other sources exhibit transient magnetar behavior. Moreover, the D$_{90}$ duration of 1E 1841-045 is the longest with nearly 160 days while those of transient magnetars are typically shorter than a month and can be as short as hours.

\begin{table}[!htbp]
    \centering
    \caption{Duration of Burst Active Episodes}
    \label{tab:burst_active_dur}
    \begin{tabular}{cccccc}
        \hline\hline
    Source & Start Time & End Time & Span & Number of bursts & D$_{90}$ \\ 
        & (UTC) & (UTC) & (days) &  & (days) \\ \hline
    SGR 0501+4516 & 080822 12:41:59 & 080903 18:53:51 & 13 & 27 & 4.72 \\ \hline
    \sgrtwo & 081003 09:02:47 & 081010 12:53:38 & 8 & 17 & 1.06 \\
            & 090122 00:53:52 & 090417 22:42:11 & 86 & 435 & 12.47 \\ \hline
    1E 1841-045 & 240820 18:39:13 & 250313 17:52:11 & 205 & 32 & 157.5 \\ \hline
    \sgrone  & 150222 17:57:05 & 150305 18:59:19 & 12 & 12 & 5.48 \\
            & 160518 08:37:07 & 160826 14:05:52 & 101 & 55 & 59.7 \\
            & 191104 01:20:24 & 191115 20:48:41 & 12 & 13 & 8.85 \\
            & 200427 18:26:20 & 200520 21:47:07 & 24 & 106 & 0.31 \\
            & 210129 02:46:23 & 210216 22:20:40 & 19 & 14 & 11.44 \\
            & 210710 15:37:56 & 210718 12:07:12 & 9 & 15 & 6.46 \\
            & 210909 18:57:15 & 211112 07:43:28 & 65 & 98 & 15.99 \\
            & 220108 14:41:47 & 220206 05:15:46 & 30 & 82 & 7.96 \\
            & 221012 12:47:04 & 221213 06:57:11 & 63 & 135 & 28.28 \\ \hline
    PSR J1846-0258 & 200718 08:24:39 & 200801 20:11:48 & 15 & 17 & 0.29 \\ \hline

    \end{tabular}

\end{table}

Two magnetars account for a large fraction of the bursts in our sample: \sgrtwo and \sgronenos. Most of the bursts from \sgrtwo were detected during its most intense bursting episode in January 2009. In particular, 341 bursts were observed on 2009 January 22. We have calculated the effective GBM exposure\footnote{That is the duration while the source location remained above the Earth's limb but excludes the time spans of spacecraft's passages through the South Atlantic Anomaly. A list of average daily exposure of the 17 bursting magnetar is given in the Appendix \ref{app_a}.} of the 17 magnetars in this catalog. The average daily GBM exposure for \sgrtwo is about 54000 s. In other words, bursts from this magnetar could not be recorded for about 37$\%$ per day. Taking this fact into account, we estimate that the source might have emitted $\sim541$ bursts on its most active day. 

Among the nine burst active episodes of \sgronenos, there are two intense bursting epochs, during which daily burst rate exceeded 50. These are 2020 April 27 and 2022 October 12 during which there were 59 and 52 bursts recorded, respectively. Given the average daily exposure time of \sgrone with GBM was calculated to be 47942 s, we estimate that the number of bursts from \sgrone on these two days could have been as high as $\sim106$ and $\sim95$, respectively.

We note that our D$_{90}$ estimates were calculated based on the number of bursts included in our catalog, which only includes bursts found within the 300-s trigger data readout. There are likely sub-threshold magnetar bursts in the continuous TTE GBM data; however, those should comprise a small fraction of the entire burst populations and should not affect our estimates here.

\subsection{Spectral Analysis Results}

We performed time-integrated spectral analysis of 1254 bursts, using four different photon models.
In \autoref{tab:spectral}, we present the fit results for all photon models (BB, BB+BB, PL, and COMPT) that provided acceptable ``goodness of fit" based on the expected $C$-stat variance for each burst (see \autoref{sec:spec_analysis}). 
We remind that all our spectral results presented here are statistically reliable resulting in well-constrained parameters in addition to the ``goodness of fit" being within 3$\sigma$.
Most of the events found within the data readout time in our sample are not necessarily weaker events, and would have triggered GBM if they occurred individually and not during the readout time of the TRIGDAT data.

% but they did not trigger the GBM detectors simply because they occurred within the readout time during which the data were accumulated for the triggered events.

In terms of signal-to-noise ratio (SNR), only 16\% (206 out of 1254) of the events provided a total SNR of $<$ 15 in the energy range we used, which is adequate for yielding statistically reliable fit results.  Additionally, only 2.8\% (35 out of 1254) of our events are detected with $<$ 5$\sigma$ in the \textit{second} brightest detectors, meaning that they could truly be considered below the trigger threshold in 8-200 keV.

Many of the spectra are adequately described by multiple photon models. Overall, we find that the two-component BB+BB model provides acceptable fits to most (68\%) of the burst spectra, while the non-thermal COMPT model sufficiently fits 46\% of the spectra in our sample (see \autoref{tab:model-stat}). We caution that these numbers do not necessarily represent the best model describing the spectra; rather, the numbers in \autoref{tab:model-stat} show those provided statistically acceptable fits. 
In the table, the bold numbers in diagonal indicate the spectra with single acceptable models. The off-diagonal numbers are for spectra acceptably fitted by the corresponding pair of models (e.g., 9 spectra are described by both BB and BB+BB). The values in parentheses duplicate the symmetric entries. The columns labeled by three-model combinations list spectra acceptably fitted by those three models, while ``\textit{All}" and ``\textit{No}" indicate spectra fitted by all four models or by none of the tested models, respectively. The final column gives the total number (and percentage) of spectra acceptably fitted by each individual model, irrespective of whether other models also provide acceptable fits.
Note that two bright bursts (254286352.375 and 254292081.916 MET) from \sgrtwo cannot be fit by any of these models used here (included as ``No" in \autoref{tab:model-stat}). These bursts will be investigated in a follow-up paper.

\begin{table}[!htbp]
    \centering
    \caption{The number of events whose spectra yielded acceptable fits for each photon model.}
    \label{tab:model-stat}
    \begin{tabular}{c|c|c|c|c||c|c|c|c||c|c|c}
        \hline\hline
     & \multicolumn{11}{c}{Number of Spectra (1254 in total)}  \\ \cline{2-12}
     Accept.& \multicolumn{4}{c||}{\textit{One or Two Models}} & \multicolumn{4}{c||}{\textit{Three Models}} && \\ \cline{2-9}
     Model(s) & /BB & /2B & /PL & /CO & BB/2B/PL & BB/2B/CO & BB/PL/CO & 2B/PL/CO & \textit{All} & \textit{No} & Total \\ \hline
        BB & \textbf{119} & 9 & 200 & 7 & \multirow{4}{*}{224} & \multirow{4}{*}{34} & \multirow{4}{*}{57} & \multirow{4}{*}{140} & \multirow{4}{*}{127} & \multirow{4}{*}{2} & 777 (62\%) \\ \cline{1-5}\cline{12-12}
        2B & (9) & \textbf{16} & 103 & 205 & & & & &  & & 858 (68\%)  \\ \cline{1-5}\cline{12-12}
        PL & (200) & (103) & \textbf{7} & 1 & & & & &  & & 859 (69\%)  \\ \cline{1-5}\cline{12-12}
       CO & (7) & (205) & (1) & \textbf{3} & & & & &  & & 574 (46\%)  \\ \hline
    \end{tabular}
    
\tablecomments{BB: single blackbody, 2B: two blackbodies (BB+BB), PL: power law, CO: COMPT models.\\
The numbers in parenthesis under \textit{One or Two Models} are duplicates of the off-diagonal entries.}
\end{table}

The simpler models (BB and PL) can fit quite high number of spectra in our sample.  When we look at the spectral parameter distributions of these models, we find that the mean BB temperature ($kT$) is 12.3\,keV, with the bursts from \sgrtwo showing higher $kT$ overall with a mean of 15.5\,keV.  The same trend of \sgrtwo being spectrally harder is also seen from the PL, although only about 60\% of the bursts fit by BB and by PL overlap.  The mean PL index is $-1.87$.
Among the spectra fit with both single BB and COMPT, there is a very strong positive correlation between $kT$ and $E_{\rm peak}$ (Spearman rank-order correlation $r_s = 0.96$, $P < 10^{-16}$). This is expected as these temperatures represent the peaks in the spectra.  When the spectra can be fit with both PL and BB or PL and COMPT, again we observe strong positive correlations between the PL indices ($\gamma$) and temperatures, $kT$ or $E_{\rm peak}$ ($r_s = 0.94$ and $r_s = 0.97$, respectively, both with $P < 10^{-16}$).  These indicate that the PL indices can confidently represent the bursts' hardness within this energy.

\begin{longrotatetable}
\begin{deluxetable*}{lccccccccccc}
\tablecaption{Spectral fit results showing all acceptable fits for each burst.  The full list is available in the electronic version of the paper. \label{tab:spectral}}
\tablewidth{0pt}
\tabletypesize{\footnotesize}
\tablehead{
\colhead{Burst} &
\multicolumn{2}{c}{BB} &
\multicolumn{3}{c}{BB+BB} &
\multicolumn{2}{c}{PL} &
\multicolumn{3}{c}{COMPT} &
\colhead{Flux} \\
\cline{2-3} \cline{4-6} \cline{7-8} \cline{9-11}
\colhead{Start Time} &
\colhead{$kT$} & \colhead{$C_{\rm stat}$/dof} &
\colhead{$kT_{\rm Low}$} & \colhead{$kT_{\rm High}$} & \colhead{$C_{\rm stat}$/dof} &
\colhead{$\gamma$} & \colhead{$C_{\rm stat}$/dof} &
\colhead{$\alpha$} & \colhead{$E_{\rm peak}$} & \colhead{$C_{\rm stat}$/dof} &
\colhead{$F$} \\
\colhead{(MET)} &
\colhead{(keV)} & \colhead{} &
\colhead{(keV)} & \colhead{(keV)} & \colhead{} &
\colhead{} & \colhead{} &
\colhead{} & \colhead{(keV)} & \colhead{} &
\colhead{($10^{-7}$ erg cm$^{-2}$ s$^{-1}$)}
}
\startdata
241101720.434 & $14.10^{+1.07}_{-0.98}$ & $118.55/130$ & \nodata & \nodata & \nodata & $-1.47^{+0.08}_{-0.09}$ & $99.17/130$ & \nodata & \nodata & \nodata & $10.20^{+0.91}_{-0.90}$ \\
241140778.728 & $13.38^{+1.84}_{-1.48}$ & $28.24/33$ & \nodata & \nodata & \nodata & $-1.41^{+0.17}_{-0.18}$ & $23.24/33$ & \nodata & \nodata & \nodata & $37.95^{+10.84}_{-5.68}$ \\
241144093.533 & \nodata & \nodata & $5.08^{+0.46}_{-0.40}$ & $20.66^{+2.13}_{-1.81}$ & $98.65/96$ & $-1.98^{+0.05}_{-0.05}$ & $110.60/99$ & $-1.48^{+0.19}_{-0.18}$ & $45.39^{+6.71}_{-5.38}$ & $101.43/98$ & $33.41^{+1.74}_{-1.64}$ \\
241150299.965 & \nodata & \nodata & $4.56^{+0.40}_{-0.35}$ & $16.79^{+1.12}_{-1.02}$ & $103.75/103$ & $-2.00^{+0.04}_{-0.04}$ & $133.28/106$ & $-1.17^{+0.17}_{-0.17}$ & $45.63^{+3.15}_{-2.85}$ & $98.76/105$ & $10.88^{+0.40}_{-0.39}$ \\
241157422.756 & $14.97^{+1.06}_{-0.94}$ & $125.05/139$ & \nodata & \nodata & \nodata & $-1.60^{+0.07}_{-0.07}$ & $100.38/139$ & $-1.00^{+0.33}_{-0.30}$ & $99.31^{+50.85}_{-20.09}$ & $95.24/138$ & $5.65^{+0.41}_{-0.39}$ \\
241163795.052 & $14.15^{+0.81}_{-0.75}$ & $56.02/91$ & \nodata & \nodata & \nodata & $-1.59^{+0.07}_{-0.07}$ & $73.74/91$ & $-0.02^{+0.42}_{-0.37}$ & $64.70^{+7.64}_{-5.75}$ & $46.12/90$ & $7.22^{+0.50}_{-0.47}$ \\
241167673.709 & $12.87^{+0.50}_{-0.47}$ & $154.57/146$ & $10.18^{+0.85}_{-0.97}$ & $26.83^{+9.18}_{-5.07}$ & $124.25/142$ & $-1.72^{+0.05}_{-0.05}$ & $182.50/146$ & $-0.23^{+0.29}_{-0.26}$ & $60.05^{+4.76}_{-3.94}$ & $132.17/145$ & $13.74^{+0.76}_{-0.74}$ \\
241167866.188 & \nodata & \nodata & $4.10^{+0.46}_{-0.39}$ & $24.67^{+4.39}_{-3.50}$ & $80.78/123$ & $-1.92^{+0.08}_{-0.08}$ & $72.12/127$ & \nodata & \nodata & \nodata & $21.06^{+1.87}_{-1.68}$ \\
241169975.910 & $9.75^{+0.40}_{-0.39}$ & $157.68/142$ & $5.07^{+0.58}_{-0.47}$ & $19.77^{+2.37}_{-1.95}$ & $80.46/138$ & $-1.95^{+0.06}_{-0.06}$ & $92.38/142$ & $-1.32^{+0.23}_{-0.22}$ & $47.91^{+6.96}_{-5.17}$ & $81.91/141$ & $16.49^{+0.96}_{-0.92}$ \\
241170949.287 & \nodata & \nodata & $5.47^{+0.33}_{-0.31}$ & $15.72^{+1.34}_{-1.12}$ & $140.02/144$ & \nodata & \nodata & $-0.81^{+0.16}_{-0.16}$ & $33.07^{+1.18}_{-1.21}$ & $148.52/147$ & $20.53^{+0.55}_{-0.53}$ \\
241173005.198 & $9.59^{+0.74}_{-0.70}$ & $71.51/81$ & \nodata & \nodata & \nodata & \nodata & \nodata & \nodata & \nodata & \nodata & $13.44^{+1.38}_{-1.24}$ \\
241179497.452 & $13.71^{+0.63}_{-0.59}$ & $169.09/148$ & $3.88^{+0.52}_{-0.46}$ & $20.23^{+1.72}_{-1.54}$ & $104.82/144$ & $-1.72^{+0.06}_{-0.06}$ & $117.30/148$ & $-1.11^{+0.24}_{-0.23}$ & $81.83^{+21.93}_{-11.84}$ & $107.99/147$ & $13.41^{+0.82}_{-0.79}$ \\
241183655.975 & \nodata & \nodata & $5.56^{+0.21}_{-0.22}$ & $12.69^{+0.24}_{-0.25}$ & $121.09/150$ & \nodata & \nodata & $-0.02^{+0.10}_{-0.10}$ & $36.50^{+0.47}_{-0.47}$ & $147.03/153$ & $102.47^{+1.75}_{-1.50}$ \\
  ...
\enddata
\tablecomments{\\
Errors are indicated in 1$\sigma$ confidence level. \\
Fluxes are reported in the 8--200 keV range with the priority BB+BB $>$ BB $>$ COMPT $>$ PL.  \\
``\nodata'' indicates the parameter was unconstrained or the model was not accepted. \\
*~ Saturated events; The saturated parts of these events were excluded from the spectral analysis.\\
 $^\dagger$ None of the photon models we use provides an adequate fit to the bursts starts with 254286352.375 MET (bn090122129) and 254292081.916 MET.
}
\end{deluxetable*}
\end{longrotatetable}

\subsubsection{Preferred Model Selections}
As can be seen from \autoref{tab:model-stat} and \autoref{tab:spectral}, more than 88\% of the spectra were adequately fit by multiple photon models.  This by itself is a useful piece of information and something to keep in mind when performing a spectral analysis of SGR bursts; however, we can also compare the model fit results to discern the statistically preferred model for each spectrum.  This enables more meaningful interpretations of the derived spectral properties in subsequent population studies.  To this end, we calculated Bayesian Information Criteria (BIC = $-2\ln\mathcal{L}_{\rm max} + k\ln N$) using the $C$-stat values obtained from the fits since $C$-stat = $-2\ln\mathcal{L}_{\rm max}$; here, $\mathcal{L}_{\rm max}$ is the maximum likelihood, $k$ is the number of free parameters in the model and $N$ is the number of data points. 
%\mg{from Eric Burns:  BIC cannot be used to discriminate BB vs BB+BB. The two options are not identifiable, i.e., if the amplitude of the second BB is 0 then the two spectral fits can be identical, so the underlying assumptions of BIC are not valid. You can read details here https://arxiv.org/abs/2410.00286. I do not think you need to implement the full simulation formalism, but a comment that this is using an approximation may be worth adding}
The difference in BIC between two spectral fits can then represent the Bayes factor for the posterior probability comparison.  
To be precise, such likelihood ratio tests assume $\chi^2$ statistics distribution. This requires the two models in comparison to be ``identifiable", meaning that different values of the parameters specify distinct models \citep{Algeri2020}.  This may not apply in the case of BB vs.~BB+BB comparison since the second BB component parameters are not known or constrained a priori \citep[e.g.,][]{Burns2024}.  Nonetheless, The BIC comparison still provide good approximations of preference, so we employ it here. 
We chose the criterion of $\Delta$BIC $\geq 10$, corresponding the Bayes factor of $\sim 148$, corresponding to $p > 1 - 1/148 = 99.3$\% by which the model with the smaller BIC is preferred \citep{KassRaftery1995}.  In case $\Delta$BIC $<$ 10, we take both models as preferred models for the purpose of this catalog.  We excluded PL in the final preferred model comparison because PL in our case is purely empirical.  Yet, as seen in \autoref{tab:model-stat}, there are 7 events that were only fit with the PL model, and the PL is inevitably the most preferred model for those events. 
During the comparison study, we also observed that PL provides statistically better fits to a significant fraction of the spectra fitted also with BB+BB, when $kT_{\rm high} \gtrsim 20$\,keV or $kT_{\rm low} \lesssim 4$\,keV.  These correspond to the blackbody peak energy of above 60 or below 11 keV, the ranges where the count rates are relatively lower due to the nature of the source or limited effective area of the detector.
The overall statistics of the preferred models are presented in \autoref{tab:model-stat-pref}. Similar to \autoref{tab:model-stat}, the bold numbers indicate the spectra with a single preferred model. The off-diagonal entries represent the number of spectra for which the corresponding pair of models are equally preferred (values in parentheses duplicate the symmetric entries). The ``\textit{All Models}" column lists spectra for which BB, BB+BB, and COMPT are equally preferred.
%, while the ``\textit{PL Only}" and ``\textit{No Model}" columns indicate spectra for which PL alone or none of the models provides statistically acceptable fit, respectively (see \autoref{tab:model-stat}). 
The ``Total" column gives the total number (and percentage) of spectra that prefer each model.
We find that 76\% of our sample events are preferentially described by single models (the bold numbers in \autoref{tab:model-stat-pref}). 
On average, events that are better described by BB or PL-only have lower photon flux than those represented by BB+BB or COMPT, both of which require substantial counts in higher energy to be well constrained.
All the analyses presented hereafter were performed using the spectral parameters of the preferred photon models presented in \autoref{tab:model-stat-pref}.

\begin{table}[!htbp]
    \centering
    \caption{Preferred models based on the $\Delta$BIC $\geq 10$ criteria excluding PL. The bold numbers indicate the spectra with a single preferred model. Others can be fit with multiple models equally preferred.}
    \label{tab:model-stat-pref}
\begin{comment}
\begin{tabular}{c|c|c|c|c|c|c|c|c}
        \hline\hline
    & \multicolumn{7}{c}{Number of Spectra (1254 in total)} & \\ \cline{2-9}
    Preferred Model(s) & \multicolumn{3}{c|}{\textit{One or Two Models}} &  \textit{All Models} &&&& Median Flux\\ \cline{2-4}
    without PL & BB & BB+BB & COMPT & BB/BB+BB/CO & \textit{PL Only}$^\dagger$ & \textit{No Model} & Total & (photons\,cm$^{-2}$\,s$^{-1}$)\\ \hline
        BB/ & \textbf{320} & 41 & 23 & \multirow{3}{*}{2} & \multirow{3}{*}{\textbf{7}} & \multirow{3}{*}{2} & 386 [31\%] & 18.6 \\ \cline{1-4}\cline{8-9}
       BB+BB/ & (41) & \textbf{372} & 231 &  &  & & 646 [52\%] & 35.7\\ \cline{1-4}\cline{8-9}
        COMPT/ & (23) & (231) & \textbf{256} &  &  & & 512 [41\%] & 45.4\\ \hline
\end{tabular}
\tablecomments{The numbers in ( ) are duplicates of the off-diagonal entries.\\
\textbf{$^\dagger$\,Only PL provided acceptable fits to these events (see \autoref{tab:model-stat})}}
\end{comment}

\begin{tabular}{c|c|c|c|c|c|c}
        \hline\hline
    & \multicolumn{5}{c|}{Number of Spectra (1245$^*$ in total)} & \\ \cline{2-6}
    Preferred Model(s) & \multicolumn{3}{c|}{\textit{One or Two Models}} &  \textit{All Models} && Median Flux\\ \cline{2-4}
    without PL & /BB & /BB+BB & /COMPT & BB/BB+BB/CO & Total & (photons\,cm$^{-2}$\,s$^{-1}$)\\ \hline
        BB & \textbf{320} & 41 & 23 & \multirow{3}{*}{2} & 386 [31\%] & 18.6 \\ \cline{1-4}\cline{6-7}
       BB+BB & (41) & \textbf{372} & 231 &  & 646 [52\%] & 35.7\\ \cline{1-4}\cline{6-7}
        COMPT & (23) & (231) & \textbf{256} &  & 512 [41\%] & 45.4\\ \hline
\end{tabular}
\tablecomments{The numbers in ( ) are duplicates of the off-diagonal entries\\
$^*$\,Excluding 9 spectra for which PL alone or none of the photon models provided acceptable fits (see \autoref{tab:model-stat})}

\end{table}

We present the distributions of the spectral parameters of these preferred models in \autoref{fig:bb_pl}, \autoref{fig:compt}, and \autoref{fig:kTdists}.
Since 91\% of the 1254 bursts in the catalog originate from four sources: \sgronenos, \sgrtwonos, SGR\,1806--20, and 1E\,1841--045 (see \autoref{tab:bursts_num_energetics}), we additionally show the smoothed parameter distributions of the bursts from these four sources in these figures, to clearly show their contributions to the overall parameter distributions.
From these distributions, we observe the following for each model:\\

%%%%%%%%%%%%%%%%%%%%%%%%%%%%%%%%%%%%%%%%%%%%%%%%%
%%%%%%%%%%%%%%%%%%%%%%%%%%%%%%%%%%%%%%%%%%%%%%%%%
%%%%%%%%%%%%%%%%%%%%%%%%%%%%%%%%%%%%%%%%%%%%%%%%%
%%%%%%%%%%%%%%%%%%%%%%%%%%%%%%%%%%%%%%%%%%%%%%%%%
\begin{figure}[htbp!]
    \centering
    \epsscale{1.15}
    \includegraphics[width=0.75 \textwidth, trim=80 162 80 340, clip]{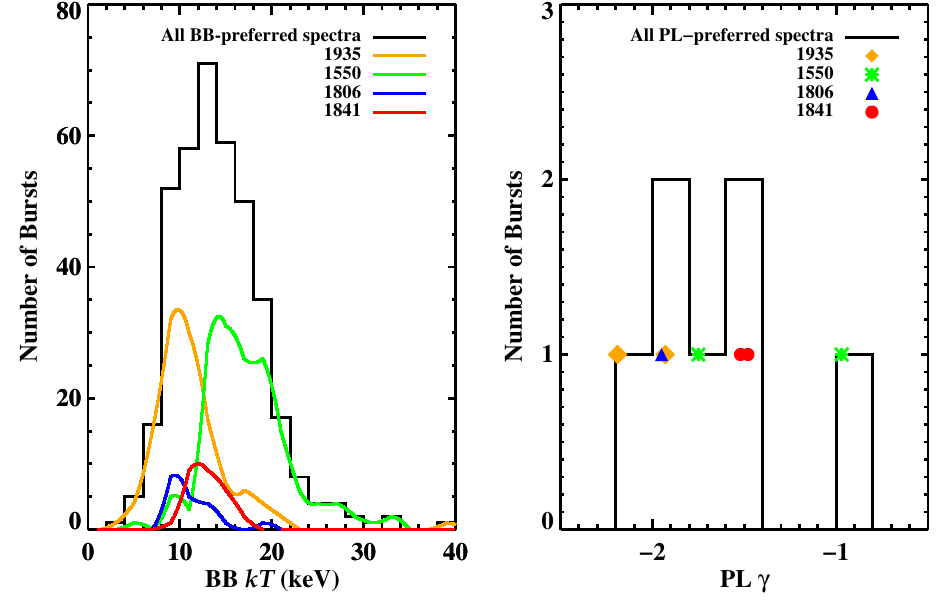}
       \caption{\textbf{[Left]} The distribution of $kT$ parameter of 386 spectra that prefer BB model is shown in black. The same distribution profiles for the most burst prolific magnetars in our sample, \sgrone (N = 119), \sgrtwo (N = 154), SGR 1806$-$20 (N = 19), and 1E 1841$-$045 (N = 27), are shown on both panels in orange, green, blue, and red, respectively.  The source specific histograms are smoothed only for clearer presentation. \textbf{[Right]} The distribution of $\gamma$ parameter of 7 spectra that favors PL model is shown in black. }
    \label{fig:bb_pl}
\end{figure}

%%%%%%%%%%%%%%%%%%%%%%%%%%%%%%%%%%%%%%%%%%%%%%%%%
%%%%%%%%%%%%%%%%%%%%%%%%%%%%%%%%%%%%%%%%%%%%%%%%%

\textbf{BB:} The mean BB temperature ($kT$) is 14.3\,keV, with the bursts from \sgrtwo showing higher $kT$ overall with a mean of 17.3\,keV (\autoref{fig:bb_pl}, left panel).
All spectrally-hard events with $kT > 25$\,keV are from \sgrtwo in the 2009 January active episode, except one event found to be from \sgrone in its 2021 September active episode (highest BB $kT$ of 38.6\,keV).\\

 \begin{figure}[htbp!]
     \centering
     \epsscale{1.15}
     \includegraphics[width=0.75 \textwidth, trim=80 162 75 340, clip]{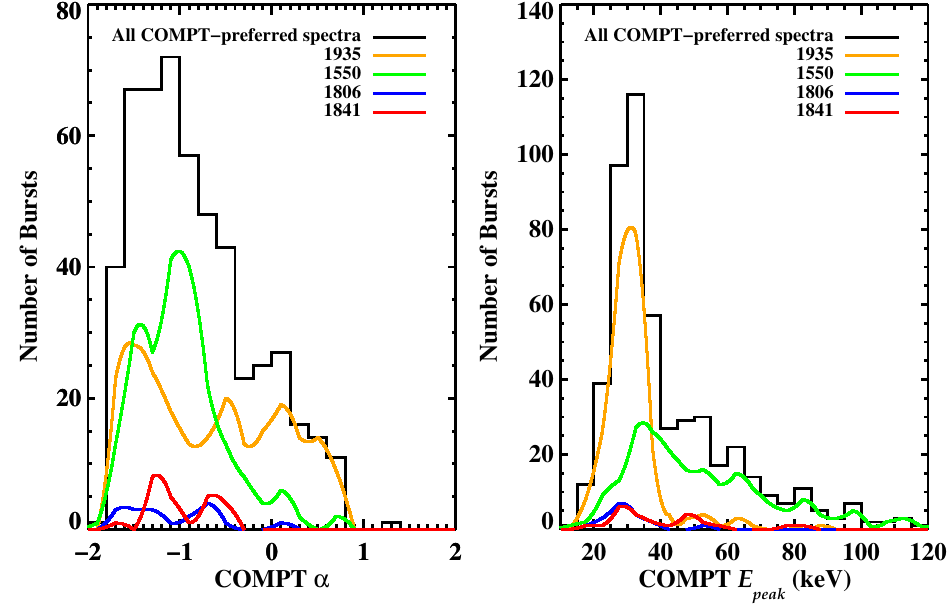}
        \caption{The distributions of $\alpha$ \textbf{[Left]} and $E_{\rm peak}$ \textbf{[Right]} parameters of 512 spectra that favors COMPT model are shown in black. The same distribution profiles for the most burst prolific magnetars in our sample, \sgrone (N = 224), \sgrtwo (N = 209), SGR 1806$-$20 (N = 18), and 1E 1841$-$045 (N = 24), are shown on both panels in orange, green, blue, and red, respectively.}
     \label{fig:compt}
 \end{figure}

\textbf{COMPT:} The photon index ($\alpha$) shows a broad distribution spanning from $-2$ to $+1$ with the mean of $-0.83$ (\autoref{fig:compt}, left panel).  No significant difference in $\alpha$ distributions are seen among the four sources although \sgrone bursts tend to have a slightly broader spread while \sgrtwo bursts peak around $-1$.  On the other hand, $E_{\rm peak}$ distributions show variations among the four sources, with the mean value of 42.1\,keV (\autoref{fig:compt}, right panel). While \sgrone bursts strongly peak at $\sim\,30$ keV, \sgrtwo bursts show a broader distribution. We observe that the spectra of \sgrtwo show higher values of $E_{\rm peak}$ than the other three sources on average, with the mean of 51.0\,keV.  This is consistent with what we observe for BB $kT$ (\autoref{fig:bb_pl}, left panel).  Also similar to the BB cases, out of 19 events with high peak energy of $E_{\rm peak}$ larger than 90 keV, 16 are from \sgrtwo in its 2008 October and 2009 January active episodes.  The other three hard events are from SGR 0501+4516 (2008), 4U 0142+61 (2015), and CXOU J164710.2-455216 (2018).\\

\textbf{BB+BB:} We found the mean temperatures of $kT_{\rm low} = 5.1$\,keV and $kT_{\rm high} = 22.5$\,keV for all 646 spectra that prefer the BB+BB model.  We present the distributions of $kT_{\rm low}$ and $kT_{\rm high}$ for \sgronenos, \sgrtwonos, SGR 1806$-$20 and 1E 1841$-$045 separately in Figure \ref{fig:kTdists}. 
It is interesting to note that unlike the cases of single-component model parameters (BB or PL), the distributions of both $kT_{\rm low}$ and $kT_{\rm high}$ are consistent for all sources. 
The $kT_{\rm high}$ distribution is wider than that of $kT_{\rm low}$; the ratios of the standard deviation ($\sigma$) and the mean value ($\mu$) of the Gaussian fits (for \sgrone and \sgrtwonos) of the $kT_{\rm low}$ and $kT_{\rm high}$ distributions are $\sigma/\mu = $ 0.21 and 0.44 for \sgrone and 0.16 and 0.36 for \sgrtwonos.
%however, the ratios of the standard deviation ($\sigma$) and the mean value ($\mu$) of the Gaussian fits (for \sgrone and \sgrtwonos) are all consistent at $\sigma/\mu \approx 6\%$ except the $kT_{\rm low}$ of \sgrtwo at 8.2\%.

%%%%%%%%%%%%%%%%%%%%%%%%%%%%%%%%%%%%%%%%%%%%%%%%%

Finally, we found no significant correlations among any pairs of the spectral parameters of the preferred models (\autoref{tab:model-stat-pref}) or with the energy flux or photon flux values. 
This is dissimilar to the \textit{NICER} catalog finding, where the spectral analysis was performed in a very limited energy range of 0.5--8\,keV \citep{Chu2026}.

\begin{figure}[htbp!]
    \centering
    \includegraphics[width=0.49\linewidth]{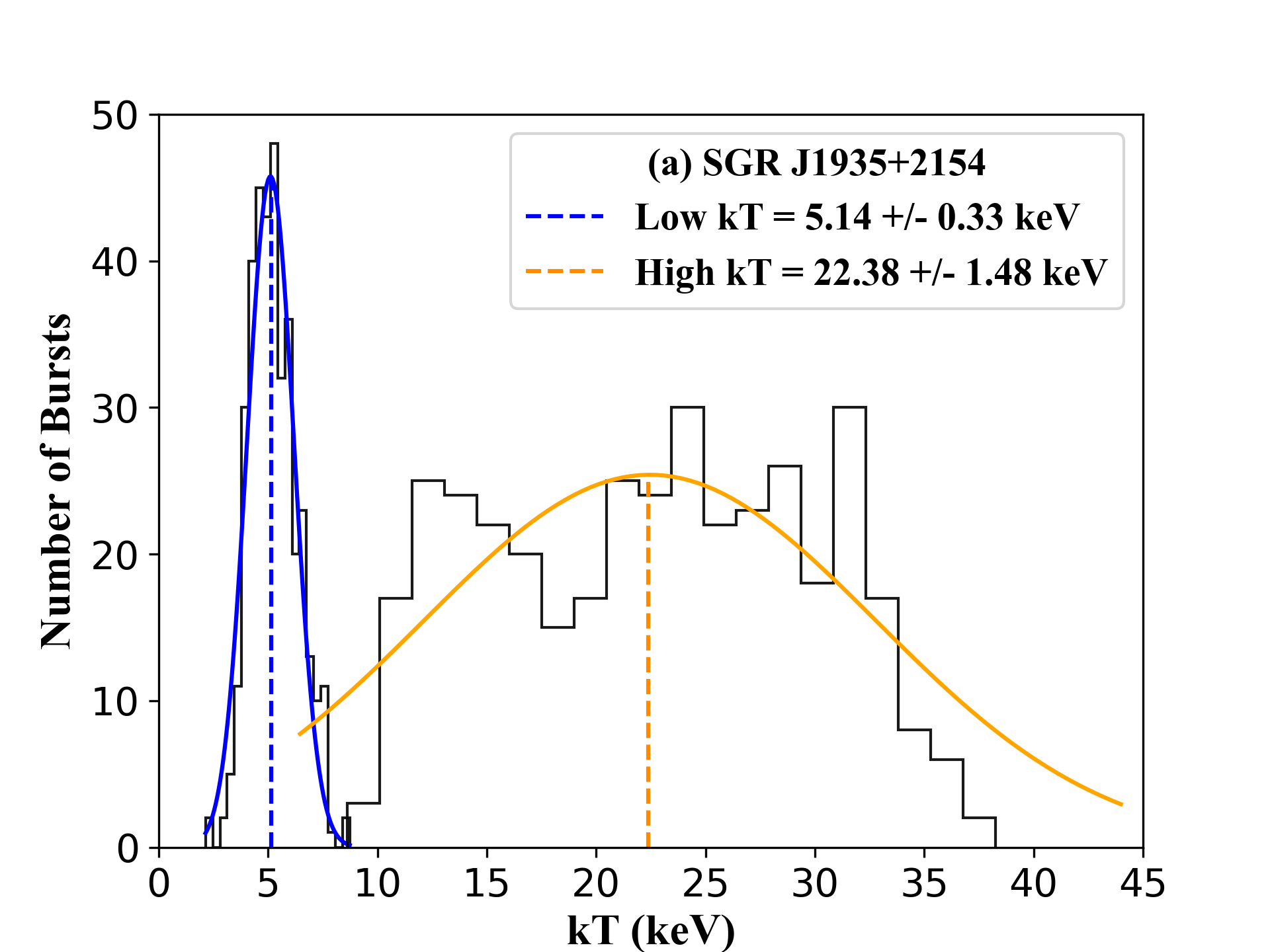}
    \includegraphics[width=0.49\linewidth]{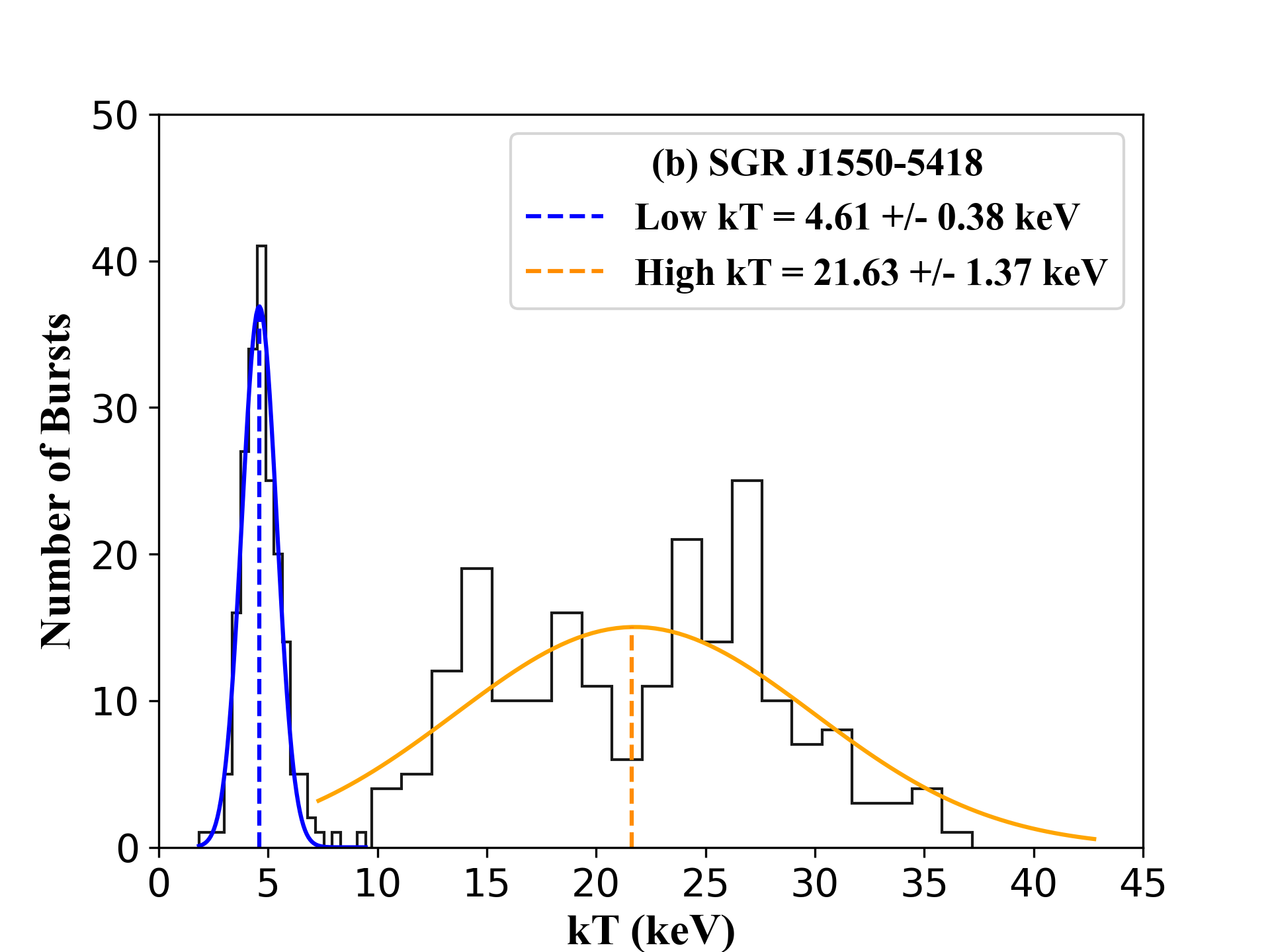} \\
    \vspace{0.2cm}
    \includegraphics[width=0.49\linewidth]{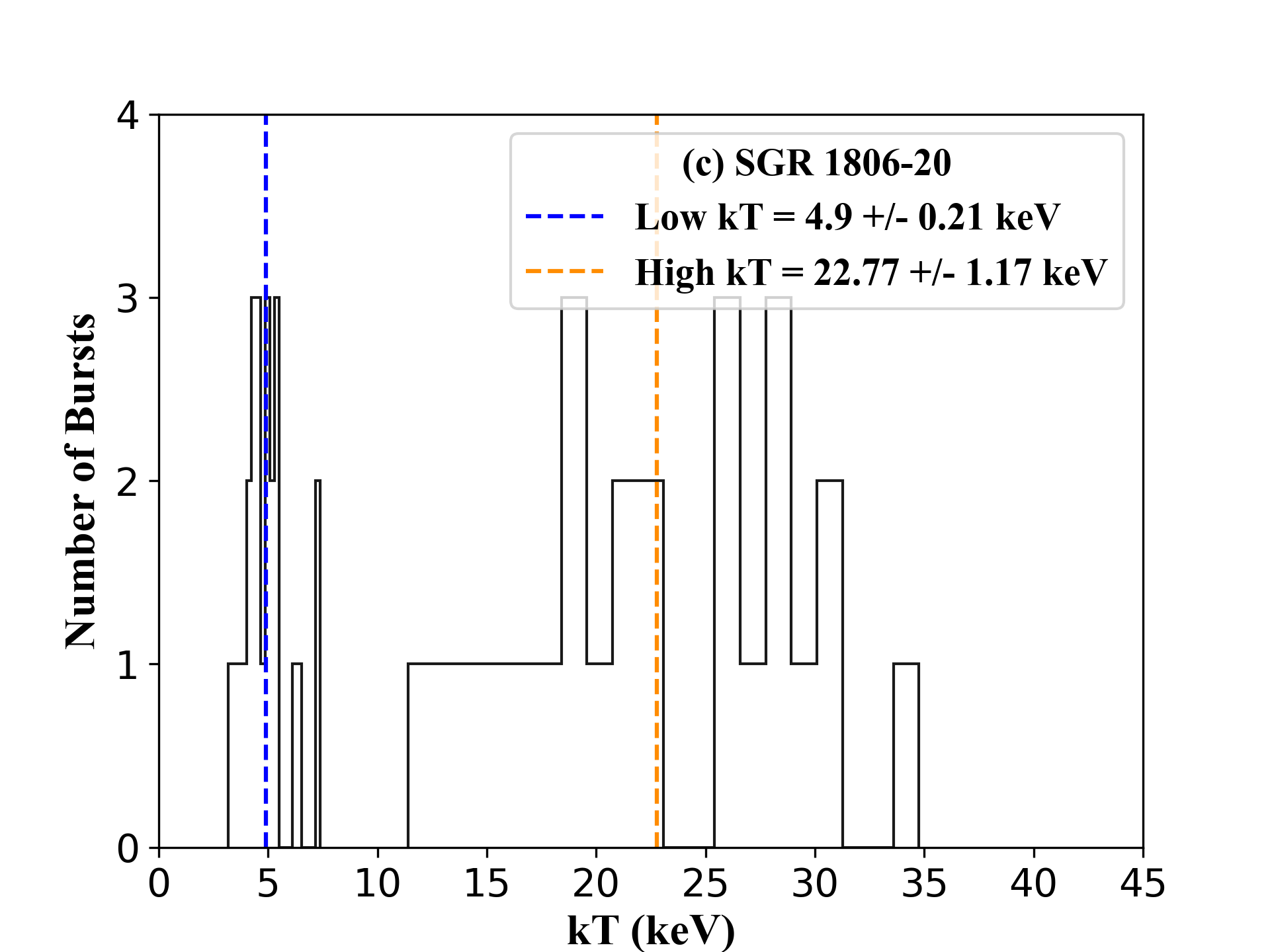}
    \includegraphics[width=0.49\linewidth]{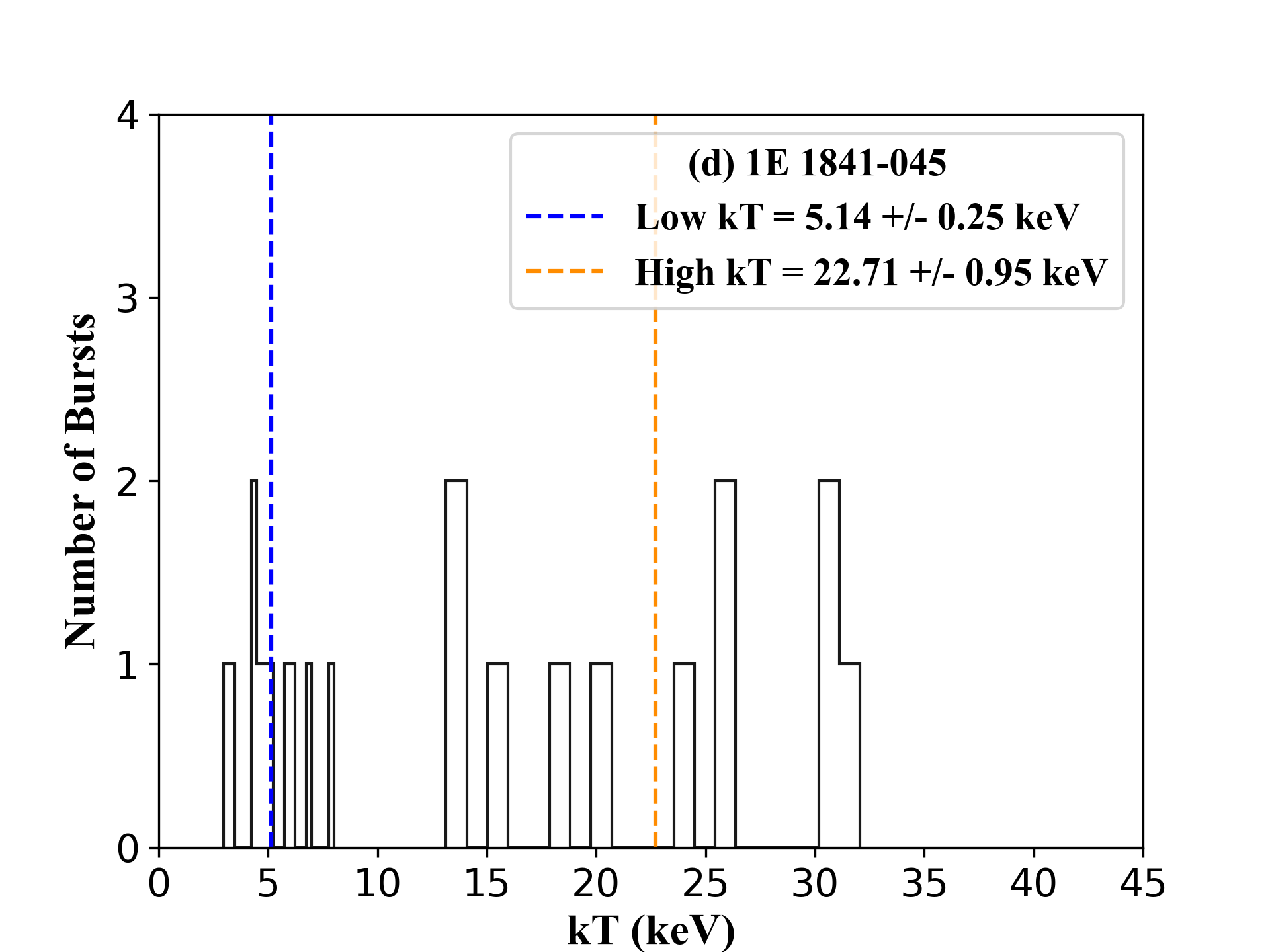} \\
    \caption{The low and high \textit{kT} distributions for the spectra that prefer BB+BB model.  Each prolific magnetar is shown separately in (a) \sgrone ($N=375$), (b) SGR J1550-5418 ($N=200$), (c) SGR 1806-20 ($N=25$), and (d) 1E 1841-045 ($N=11$).  For (a) and (b), the Gaussian fits and means were shown with the curves and the dashed lines. For (c) and (d), the weighted mean is shown instead due to poor gaussian fits of low counts.}
    \label{fig:kTdists}
\end{figure}

\section{Discussion}\label{sec:disc}

\fermi provided the longest, highly sensitive, uninterrupted monitoring of the magnetar bursts with unprecedented level of spectral and temporal data resolution. As a result, it played a crucial role in the discovery of four new magnetars and recorded over thousand bursts with high quality during the reactivation of 13 known magnetars. Our GBM 17-year catalog contains 1254 typical magnetar bursts from 17 galactic magnetars. Below, we discuss general collective characteristics of these bursts between different magnetar sources, as well as burst properties of different burst active episodes of the same magnetar. We also compare and emphasize our results with the results of 5-year GBM magnetar burst catalog \citep{Collazzi2015}, as well as recently released magnetar burst catalogs compiled using observations with \textit{NICER} \citep{Chu2026} and INTEGRAL/IBIS \citep{Pacholski2026}.

\subsection{Burst Clustering Behavior and D$_{90}$ Durations} \label{clustering}

Magnetars display diverse burst clustering behavior. Eight out of 17 sources in our sample emitted bursts in clustered manner, namely SGR 0501+4516, \sgrtwonos, SGR 1806$-$20, 1E 1841$-$045, \sgronenos, SGR 1830$-$0645, SGR J1555.2$-$5402 and PSR 1846$-$0258. Among those eight, bursts from five magnetars fall into the category of higher level of bursting at the onset of burst active episode. There are a few facts to note in that respect. \sgrone is by far the most frequently recurring magnetar, as already suggested by \citet{lin2020a}. The D$_{90}$ duration of its burst active episodes varies between 0.3 days and two months but typically between 5 to 15 days (median value of 8.85 days). 1E 1841-045 exhibits the longest D$_{90}$ burst active epoch duration with 157.5 days while that of PSR J1846 is the shortest, 0.29 days. The duration of burst active episodes of the other two sources, SGR 0501+4516 and \sgrtwo are similar to the active duration of \sgronenos. Two magnetars worth mentioning at this point: SGR 1806-20, the source that emitted a giant flare in December 2004, has shown densely clustered bursting episodes in 1996, 1998 \citep{Gogus01} and in 2003--2004 \citep{Pacholski2026}. It has been occasionally emitting bursts in much less clustered manner in the time span of interest here. The other magnetar to mention is SGR 1900+14, the source of the first galactic magnetar giant flare in August 1998. After the giant flare, it became burst active in 2001 and 2006 \citep{Israel2008}. It has not emitted any bursts since 2007 November 26, including nearly the entire time span of \ferminosp.
Both of these sources were previously found to have long D$_{90}$ between $\sim$90--300\,days before (and after, in case of SGR 1900+14) the respective giant flares \citep{Gogus2014}, similar to D$_{90}$ of 1E 1841-04 in our sample.

\subsection{Burst Energetics}

The amount of energy released with bursts is an important parameter that could bring to light the internal mechanism leading to these events. We calculated the burst energies assuming isotropic emission and using the most accurate distance measurement ($d$) currently reported in the literature. In \autoref{tab:bursts_num_energetics}, we list the extreme energetics for each source along with the total and average energies. It is important to note that the distance measurement has a significant contribution to energetics calculations ($E \propto d^2$) and any improvement in distance measurements could change the burst energy estimations. 

Overall, the energy released isotropically with magnetar bursts in our sample range between 2$\times10^{37}$ and 4$\times10^{41}$ erg. Here, the lowest energy bound can be used to infer minimum conditions to ignite magnetar burst events. We find that the bursts from Swift J1822.3$-$1606, which is the second magnetar with low inferred dipolar field strength, correspond to the lowest average burst energetics of 2.3$\times10^{37}$ erg. It has been widely suggested that even if the dipolar magnetic field strength is of the order of $10^{12}$ G, highly complex higher multipoles can possess much stronger magnetic field strength that is capable of perturbing the solid neutron star crust and give rise to bursts. It is also interesting to note that the energies of the two bursts observed from 1E 2259+586, which is a persistent magnetar with the lowest inferred dipolar field strength of 6$\times10^{13}$ G, are on the low burst energy side.

The bursts from the four most prolific sources of our sample, namely \sgronenos, \sgrtwonos, SGR 1806$-$20 and 1E 1841$-$045, along with SGR J1555.2$-$5402 and SGR J1745$-$2900 correspond to average burst energies above $10^{39}$ erg, while bursts from other sources emitted less than 6.3$\times10^{38}$ erg of energy per burst on average. 
It is worth noting that 1E 1841$-$045 is the brightest persistent source (i.e., persistently emitting in soft and hard X-rays).  However, unlike 1E 1841$-$045, the burst energies from the second and third brightest persistent magnetars, 4U 0142+61 and 1RXS J170849.0$-$400910, are among the low energy group emitting 1.4$\times10^{38}$ erg and 2.2$\times10^{38}$ erg per burst, respectively.

\begin{deluxetable}{lcrrrrc}[h!]
\tablecaption{Energetics estimates for each source. The minimum and maximum energy emitted by the dimmest and brightest burst ($E_{min}$, $E_{max}$), the total emitted energy from all bursts ($E_{total}$), and the average emitted energy per burst ($E_{avg}$) are calculated using the distance ($d$).
\label{tab:bursts_num_energetics}}
\tablehead{
\colhead{\bf Source} & \colhead{\bf Number of Bursts} & \colhead{\bf $E_{min}^{\dagger}$} & \colhead{\bf $E_{max}^{\dagger}$} & \colhead{\bf $E_{total}^{\dagger}$} & \colhead{\bf $E_{avg}^{\dagger}$} & \colhead{\bf Distance, $d$} \\
 &  & \colhead{\bf (10$^{38}$ erg)} & \colhead{\bf (10$^{38}$ erg)} & \colhead{\bf (10$^{38}$ erg)} & \colhead{\bf (10$^{38}$ erg)} &\colhead{\bf (kpc)}
}
\startdata
\sgrone&567&2.26&3922.26&45334.20&79.95&9.0$^{(1)}$ \\
SGR J1550$-$5418$^{\dagger\dagger}$&459&0.57&691.52&8835.32&19.25&5.0$^{(2)}$ \\
SGR 1806$-$20&57&4.56&210.16&1503.78&26.38&8.7$^{(3)}$ \\
1E 1841$-$045&56&4.99&199.49&1330.17&23.75&8.5$^{(4)}$ \\
SGR 0501+4516&27&0.34&71.60&158.00&5.85&2.0$^{(5)}$ \\
SGR J1555.2$-$5402&22&6.11&26.77&231.73&10.53&10.0$^{(6)}$ \\
PSR J1846$-$0258&17&2.38&9.02&66.72&3.92&6.0$^{(7)}$ \\
SGR 1830$-$0645&13&1.02&2.27&21.18&1.63&4.0$^{(8)}$ \\
CXOU J164710.2$-$455216&12&1.11&13.05&39.39&3.28&3.9$^{(9)}$ \\
4U 0142+61&10&0.78&3.94&21.68&2.17&3.6$^{(10)}$ \\
Swift J1822.3$-$1606&4&0.20&0.27&0.92&0.23&1.6$^{(11)}$ \\
SGR 0418+5729&2&0.63&0.94&1.57&0.79&2.0$^{(12)}$ \\
1E 1048.1$-$5937&2&5.40&7.18&12.58&6.29&9.0$^{(10)}$ \\
1RXS J170849.0$-$400910&2&1.20&1.66&2.86&1.43&3.8$^{(10)}$ \\
1E 2259+586&2&0.96&1.00&1.96&0.98&3.2$^{(13)}$ \\
SGR J1745$-$2900&1&16.12&16.12&16.12&16.12&8.3$^{(14)}$ \\
Swift J1818.0$-$1607&1&3.29&3.29&3.29&3.29&4.8$^{(15)}$
\enddata
\tablecomments{\\
$^{\dagger}$ In 8$-$200 keV. \\
$^{\dagger\dagger}$ Also known as 1E\,1547.0-5408.
\\
References for distance estimates: (1) \cite{Zhong2020}; (2) \cite{Tiengo2010}; (3) \cite{Bibby2008}; (4) \cite{Tian2008}; (5) \cite{Xu2006}; (6) \cite{Enoto2021}; (7) \cite{Leahy2008_psr1846}; (8) \cite{Younes2022_sgr_1830}; (9) \cite{Kothes2007}; (10) \cite{Durant2006}; (11) \cite{Scholz2012}; (12) \cite{van2010}; (13) \cite{Kothes2012}; (14) \cite{Bower2014}; (15) \cite{Lower2020}.
}
\end{deluxetable}

We also constructed the differential burst fluence distribution for the four most prolific magnetars in our sample, as well as for all bursts and present them in  \autoref{fig:logNlogS}. We find that these fluence distributions follow power laws with indices of 1.61$\pm$0.06 for all bursts, 1.53$\pm$0.09 for \sgrone bursts, 1.54$\pm$0.10 for \sgrtwo bursts, 1.73$\pm$0.44 for SGR 1806$-$20 bursts and 1.79$\pm$0.43 for 1E 1841$-$045 bursts. Even though there is slight steepening in the power law trend for the burst fluence distribution of the last two magnetars here, those burst samples are much smaller to make a conclusive statement. Overall, all fluence distributions studied here are consistent with one another within errors. Moreover, the power law indices obtained are consistent with that of self organized critical systems \citep{Gogus99,Gogus2000,Aschwanden25}.

\begin{figure}[htbp!]
    \centering
    \includegraphics[width=0.6 \textwidth, trim=85 95 60 265, clip]{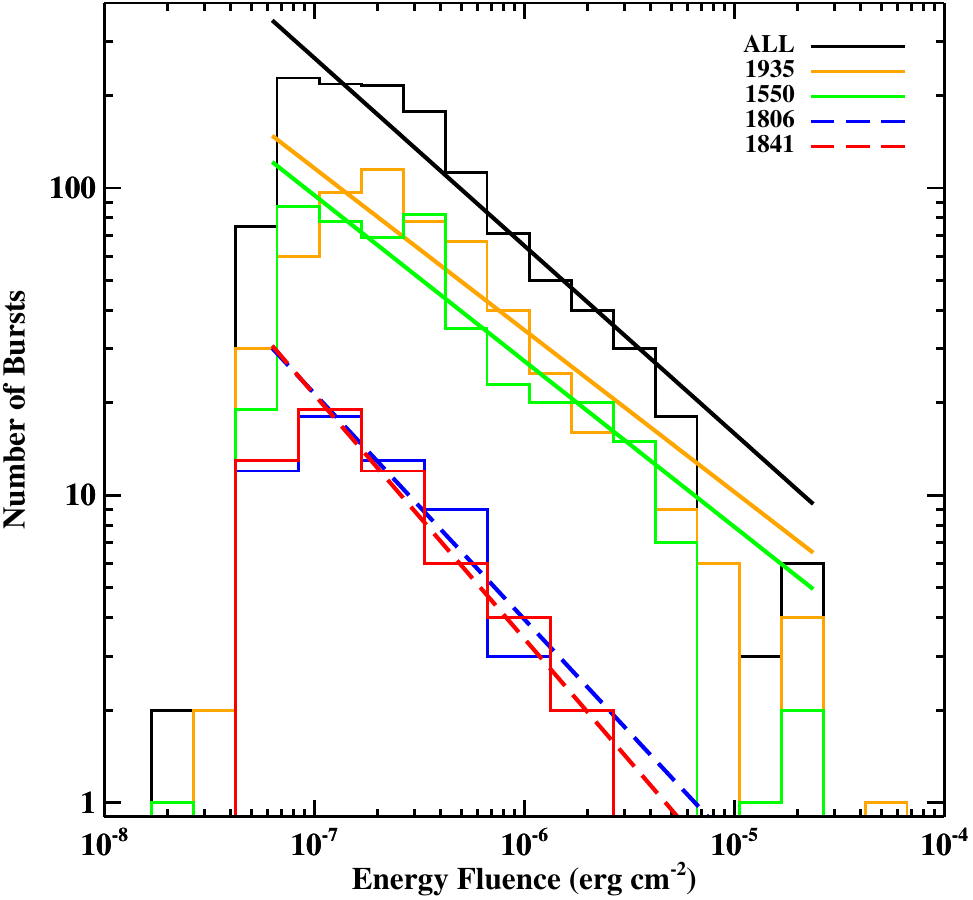}
    \caption{Differential fluence distributions of bursts from all sample (black), \sgrone (orange), SGR J1550$-$5418 (green), SGR 1806$-$20 (blue-dashed), and 1E 1841$-$045 (red-dashed). The best-fitting power-law model is overlaid on each distribution. The fluence values were calculated over the 8$-$200 keV energy range. 
    %The corresponding energy range to the fluence distribution of all bursts is $2\times10^{37} - 4\times10^{41}$ erg.
    }
    \label{fig:logNlogS}
\end{figure}

%IBIS -- Figure 6 -- (the distribution of the cumulative energy fluence), indexes are
%232 SGR 1935 bursts: 1.92 $\pm$ 0.10
%146 SGR 1550 bursts: 1.76 $\pm$ 0.04
%934 SGR 1806 bursts: 1.95 $\pm$ 0.06

\subsection{Spectral Properties}

Previously, the spectra of bursts from \sgrtwo in its two active episodes in October 2008 and January 2009 were found to display remarkably different characteristics. In the earlier epoch, burst spectra were mostly consistent with a single blackbody function \citep{vonKienlin2012}, while in the later episode, burst spectra were much harder \citep{vdhorst12}. Our systematic analysis here agree with both studies and confirm spectral evolution of bursts from an active episode to another in just few months, with the earlier episode showing more preference towards single BB and the spectra of later episode described with COMPT or BB+BB. Note the fact that most energetic bursts from \sgrtwo were emitted on 2009 January 22, during which its magnetosphere is likely loaded with highly energetic pairs, causing efficient setup for magnetospheric scattering, therefore, much harder spectra of the emerging radiation. On the other hand, we do not observe any significant variations in overall  spectral parameters of \sgrone bursts in the five active episodes investigated: average BB temperature varied from 9.4 to 12.2 keV, average BB+BB temperatures were between 4.6 and 6.1 keV for $kT_{\rm low}$ and between 22.1 and 23.1 keV for $kT_{\rm high}$. Average $E_{\rm peak}$ of the COMPT model (which is related to the magnetospheric electron temperature), on the other hand, displays marginal variation from 35.4 keV in 2015$-$2016 to 28.0 keV in 2019$-$2020 after a few-year quiescent period, and then to 32.9 keV in 2021. In the last two active episodes of 2022, we find the average $E_{\rm peak}$ value as 28.9 and 30.2 keV. 

Among the most frequently bursting sources in our sample, we find that \sgrtwo bursts are, on average, spectrally harder than those of \sgronenos. This is evident for bursts whose spectra are best represented with BB and COMPT models. We find the average BB temperature for \sgrtwo bursts as 17.3 keV while that of \sgrone bursts as 11.2 keV. Similarly, the average $E_{\rm peak}$ of \sgrtwo is 52.1 keV while that of \sgrone is 31.5 keV. This difference is also clearly seen in the right panel of \autoref{fig:compt}. There are very few \sgrone bursts with $E_{\rm peak}$ larger than 40 keV, meanwhile bulk of $E_{\rm peak}$ values of \sgrtwo bursts are spread widely between 20 and 80 keV. This can also be considered as an indication of differences in  magnetospheric processes on the emerging radiation in these two magnetars.

The integrated spectra of 646 bursts in our sample are preferentially described with the BB+BB model. Based on the resulting blackbody temperatures and flux values (therefore, the luminosities), we computed the corresponding radius ($R$) of the spherical blackbody emitting region, given the distance to each magnetar source. In \autoref{fig:r2vskt}, we present the blackbody emitting area ($R^2$) against the corresponding blackbody temperature of bursts from the four magnetars: \sgronenos, \sgrtwonos, SGR 1806--20 and 1E 1841--045. This way, we can crudely diagnose whether the burst emission follows the Stefan-Boltzmann Law, which states that for a perfect blackbody, the total energy emitted per unit surface area is proportional to the fourth power of its temperature (i.e., $R^2$ $\propto$ $kT^{-4}$). We find that $R$$^2$ vs.\,$kT$ trends of \sgrtwonos, SGR 1806--20 and 1E 1841--045 follow power law with indices of $-$5.39$\pm$0.04, $-$4.84$\pm$0.13 and $-$5.43$\pm$0.20, respectively (see also \autoref{tab:r_kT_fit_results}), showing significant deviation from the Stefan-Boltzmann Law in the emission behavior of bursts from these three magnetars. 

\begin{figure}[h!]
    \centering
    \includegraphics[width=0.45\linewidth, trim=60 75 80 260, clip]{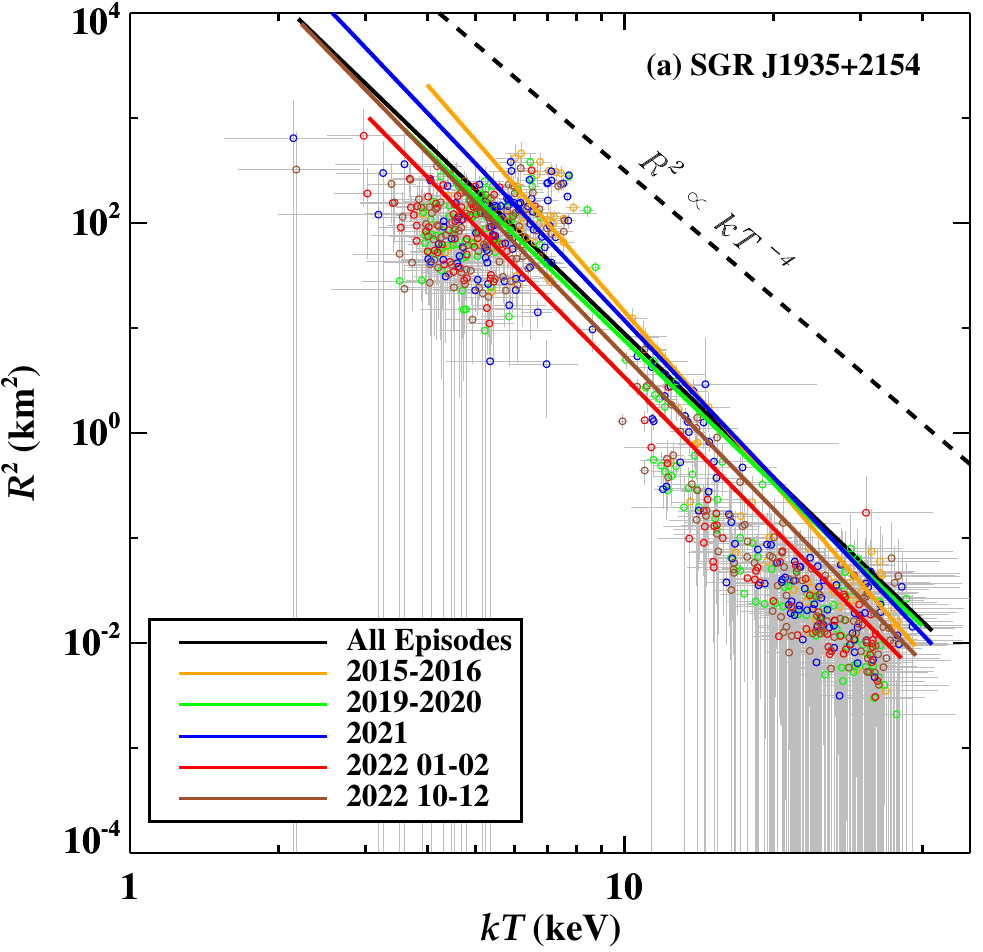}
    \includegraphics[width=0.45\linewidth, trim=60 75 80 260, clip]{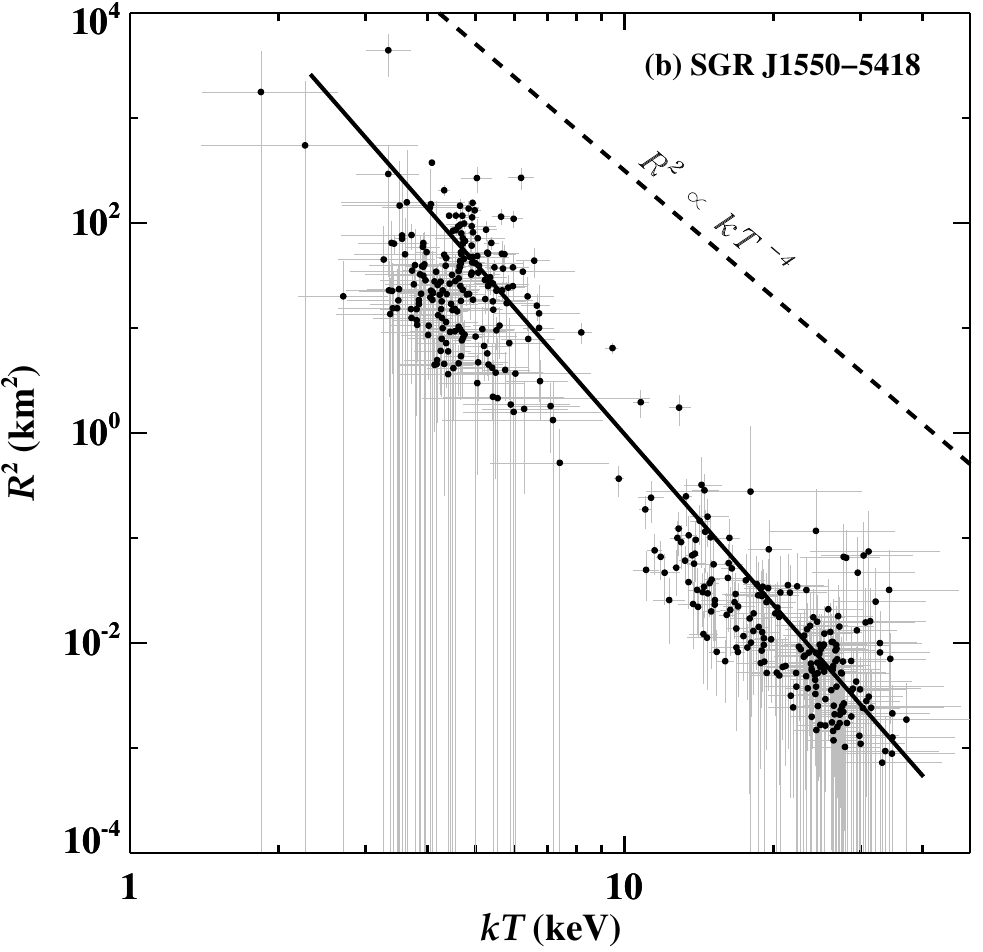} \\
    \vspace{0.2cm}
    \includegraphics[width=0.45\linewidth, trim=60 75 80 260, clip]{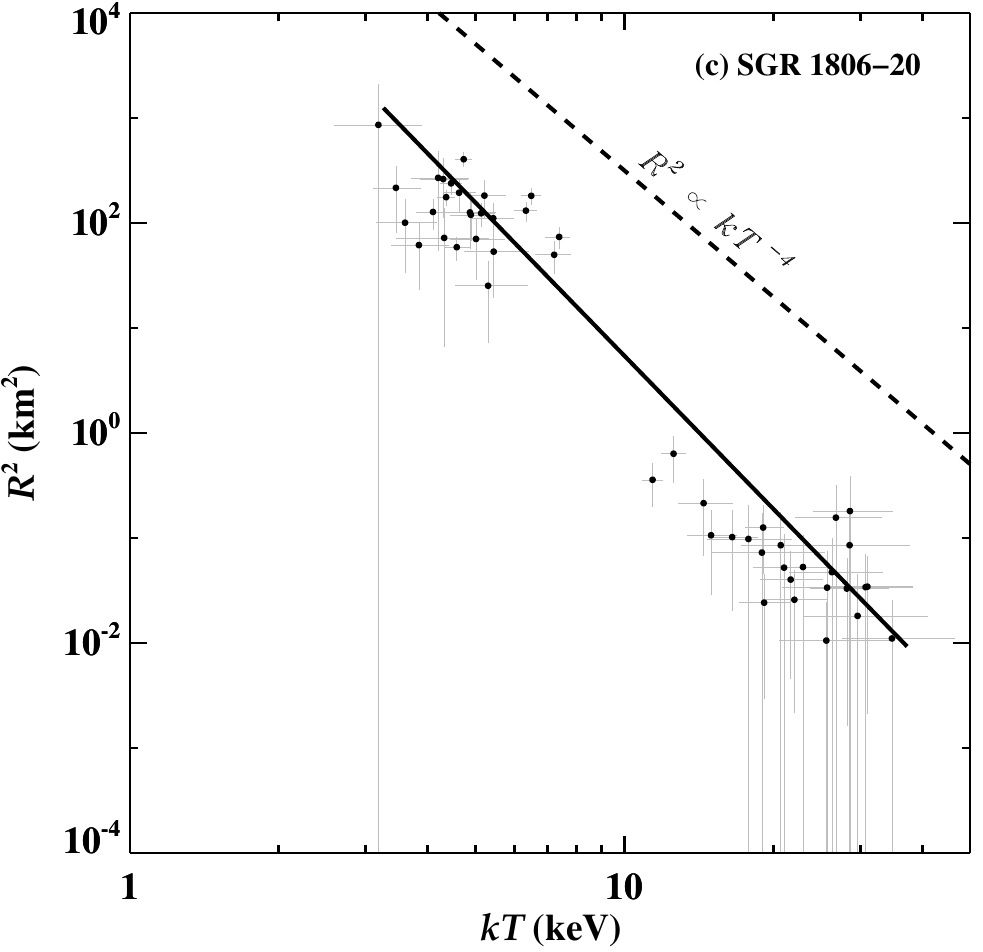}
    \includegraphics[width=0.45\linewidth, trim=60 75 80 260, clip]{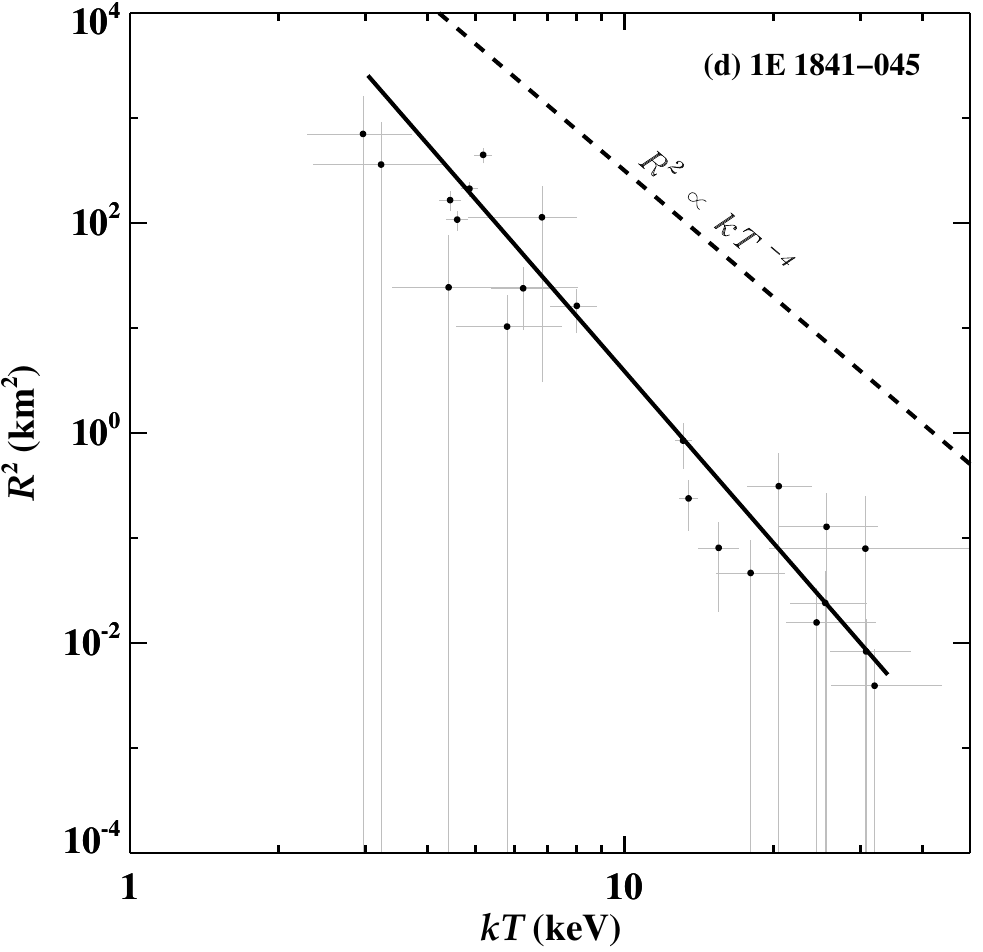} \\
    \caption{$R^2$ vs. $kT$ trends of the burst spectra that are consistent with the BB+BB model for the sources: (a) \sgronenos, (b) \sgrtwonos, (c) SGR 1806$-$20, and (d) 1E 1841$-$045. The best-fitting power-law model is overlaid on each distribution. Note that the dashed line is illustrative for $R^2 \propto kT^{-4}$ for a luminosity of 4$\times10^{43}$ erg/s.}
    \label{fig:r2vskt}
\end{figure}

In the case of \sgronenos, the $R$$^2$ vs.\,$kT$ trend of bursts spectra consistent with BB+BB can also be described with a power law, which yields an index of $-$4.55$\pm$0.03. This also corresponds to a significantly steeper trend than what is expected from a perfect blackbody case. Given numerous reactivations of \sgrone over the course of our timeline, we could also investigate whether $R$$^2$ vs.\,$kT$ trend of bursts of the same source varies from one activity episode to the other. For this purpose, we grouped active episodes of \sgrone in \autoref{tab:burst_active_dur} to contain at least 40 bursts, which resulted in grouping time periods of 2015$-$2016, 2019$-$2020, 2021, 2022 January$-$February and 2022 October$-$December. Modeling $R$$^2$ vs.\,$kT$ trends in each of these epochs with a power law yields indices ranging from $-$4.52$\pm$0.06 to $-$5.41$\pm$0.10 (see Table \ref{tab:r_kT_fit_results} for the full list of indices). In the same table, we also list the energetics of the corresponding bursts. We find no systematic relation between the power law index of $R$$^2$ vs.\,$kT$ trends with neither the total burst energy nor the average burst energy.

\begin{deluxetable}{lccccc}[h!]
\tablecaption{Indices of the power law fits to $R^2$ vs.\,$kT$ relations of bursts described with BB+BB spectra and the corresponding burst energetics: the total energy emitted per episode ($E_{total}$) and the average emitted energy per burst ($E_{average}$) using the distances shown in \autoref{tab:bursts_num_energetics}. \label{tab:r_kT_fit_results}}
\tablehead{
\colhead{\bf Source} & \colhead{\bf Active} & \colhead{\bf Index} & \colhead{\bf Number of} &\colhead{\bf $E_{total}^{\dagger}$} &\colhead{\bf $E_{average}^{\dagger}$}\\
 & \colhead{\bf Episode} & & \colhead{\bf Bursts} & \colhead{\bf (10$^{40}$ erg)} & \colhead{\bf (10$^{39}$ erg)}
}
\startdata
\multirow{7}{*}{\sgrone} & All & $-$4.55 $\pm$ 0.03 &  375 &  362.9 & 9.7  \\
 & 2015$-$2016 &  $-$5.41 $\pm$ 0.10 & 43 &  46.4 & 10.8 \\
 & 2019$-$2020 & $-$4.52 $\pm$ 0.06 & 80 &  84.4 & 10.6 \\
 & 2021 Jan$-$Jul & $-$4.96 $\pm$ 0.06 & 90 &  136.1 & 15.1 \\
 & 2022 Jan$-$Feb & $-$4.78 $\pm$ 0.08 & 60 & 27.2 & 4.5 \\
 & 2022 Oct$-$Dec & $-$4.84 $\pm$ 0.06 & 102 & 68.9 & 6.8 \\
\tableline
\sgrtwo & All & $-$5.39 $\pm$ 0.04 & 200 & 55.0 & 2.8 \\
\tableline
SGR 1806$-$20 & All & $-$4.84 $\pm$ 0.13 & 25 & 7.4 & 2.9 \\
\tableline
1E 1841$-$045 & All & $-$5.43 $\pm$ 0.20 & 11 &  4.1 & 3.7  \\
\enddata
\tablecomments{\\
$^{\dagger}$ In 8$-$200 keV.
}
\end{deluxetable}

\citet{lin2020a} reported that $R$$^2$ vs.\,$kT$ trend of \sgronenos bursts in the 2015$-$2016 active episodes follows a power law with an index of $-$4.2$\pm$0.3, that is consistent with the emission from a blackbody in thermal equilibrium. Our results for the same activity episode yield an index of $-$5.41$\pm$0.10 that is significantly apart. For the 2019$-$2020 burst epoch, \citet{lin2020b} reports an index of $-$3.6$\pm$0.2, while ours is $-$4.52$\pm$0.06. 
Note that the samples of \citet{lin2020a,lin2020b} contains 127 and 148  bursts, respectively, of which 57 and 35 untriggered bursts identified in the continuous TTE data and are therefore not included in our sample. Besides, we identified eight additional bursts in 2019$-$2020 active episode. Furthermore, whereas \citet{lin2020a,lin2020b} included all spectra that were adequately fitted by the BB+BB model, we derive the $R$$^2$ vs.\,$kT$ relation only from spectra for which BB+BB is the preferred model according to the BIC. These differences in burst samples and model-selection criteria likely account for the discrepancy between the derived power-law indices.
It is also important to note that although typical magnetar bursts are short, they show significant spectral evolution throughout the event \citep{Younes2014,Keskin2024,Demirer2025}. Therefore, even if the burst emission regions might correspond to the state of quasi-equilibrium, variations in temperature would naturally deviate the trend from a perfect blackbody. 
    
We also constructed the $R$$^2$ vs.\,$kT$ trends for the bursts from the four most prolific sources in our sample, combining events whose spectra are best represented with BB only and BB+BB only (see \autoref{fig:r2vskt_bb_2bb}). In other words, the data represent parameters obtained from distinct bursts. We find that for \sgronenos, temperatures of the BB and BB+BB models form a continuum in the $R$$^2$ vs.\,$kT$ plane, and the overall trends follow power laws with indices that are consistent with each other. In case of \sgrtwonos, bulk of the temperatures of the BB model is consistent with the lower end of the $kT$$_{\rm high}$ population of the BB+BB model, and the power law index of the $R$$^2$ vs.\,$kT$ trend for the BB temperatures is significantly different than that of BB+BB temperatures (see  \autoref{fig:r2vskt_bb_2bb}, panel b).
We observe similar situations for the bursts from SGR 1806$-$20 and 1E 1841$-$045 (see \autoref{fig:r2vskt_bb_2bb}, panel c and d). Overall, we find that the $R$$^2$ vs.\,$kT$ trends for a single blackbody model represent a closer agreement with the expectation of Stefan-Boltzmann law. Detailed investigations of the $R$$^2$ vs.\,$kT$ trends of magnetar bursts and the cause of deviation from the Stefan-Boltzmann law will be studied elsewhere.

\begin{figure}[h!]
    \centering
    \includegraphics[width=0.45\linewidth, trim=60 75 80 260, clip]{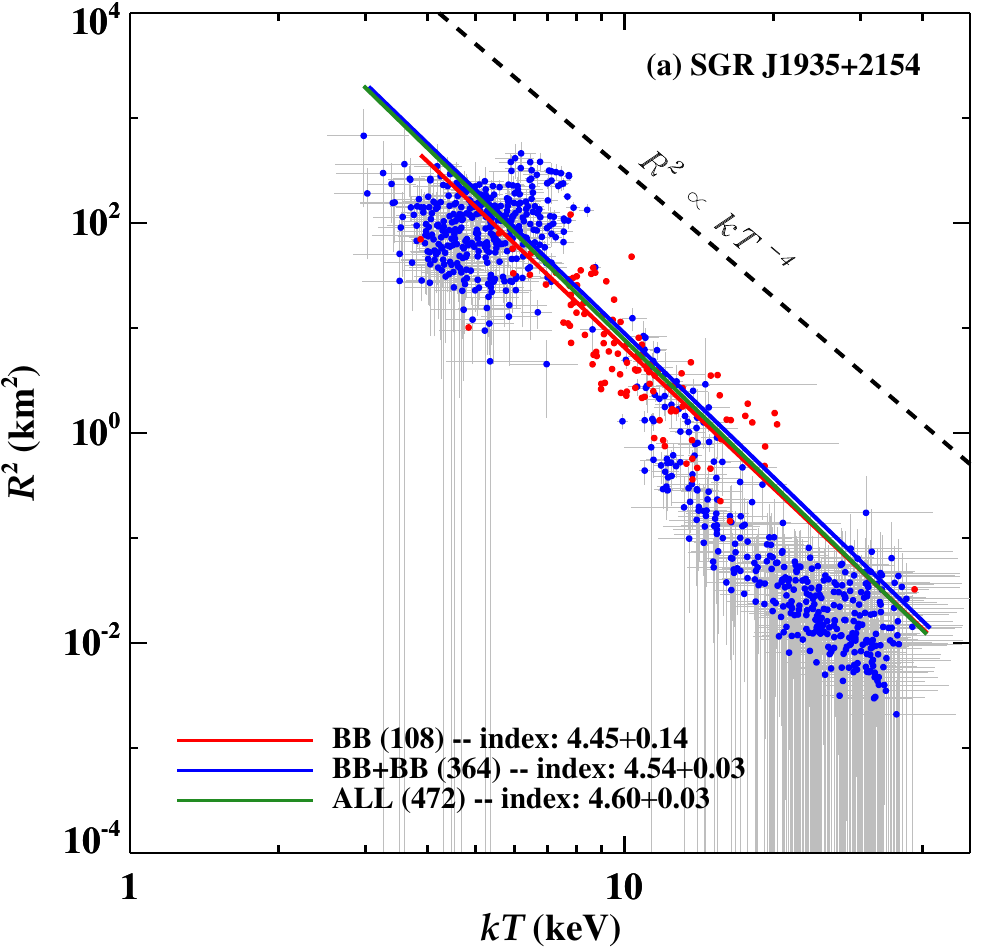}
    \includegraphics[width=0.45\linewidth, trim=60 75 80 260, clip]{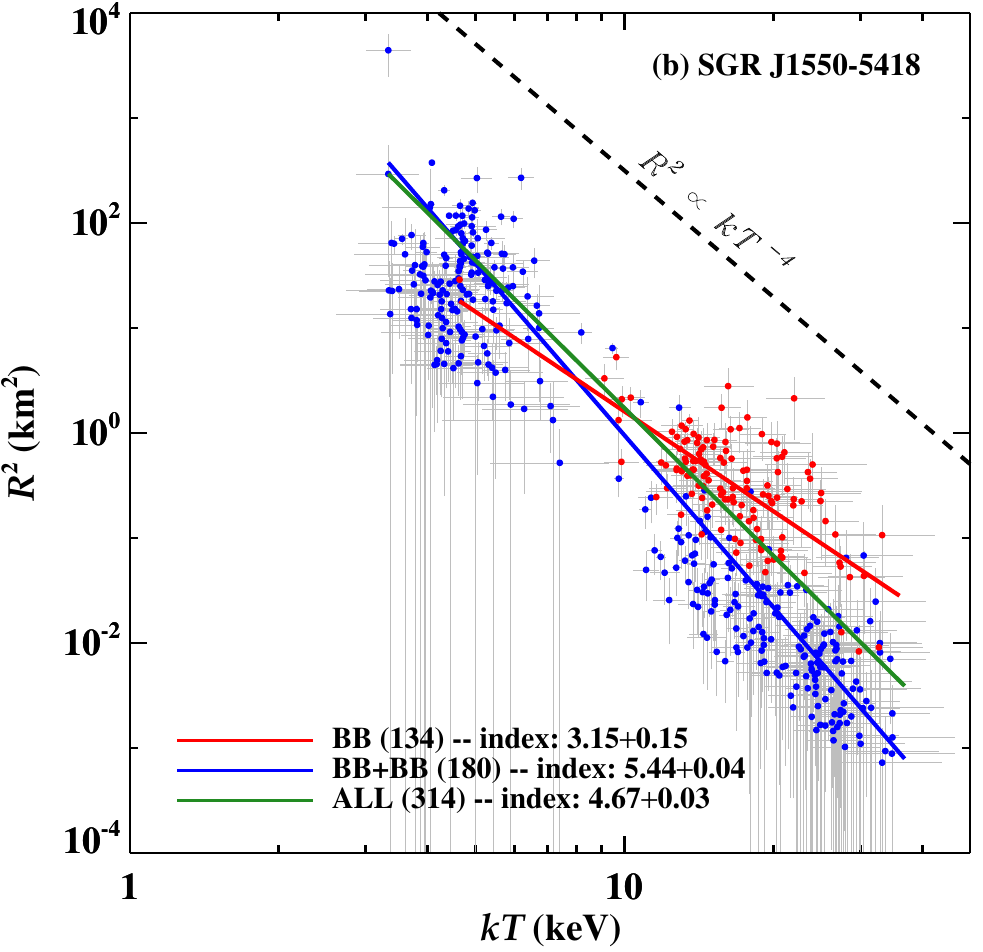} \\
    \vspace{0.2cm}
    \includegraphics[width=0.45\linewidth, trim=60 75 80 260, clip]{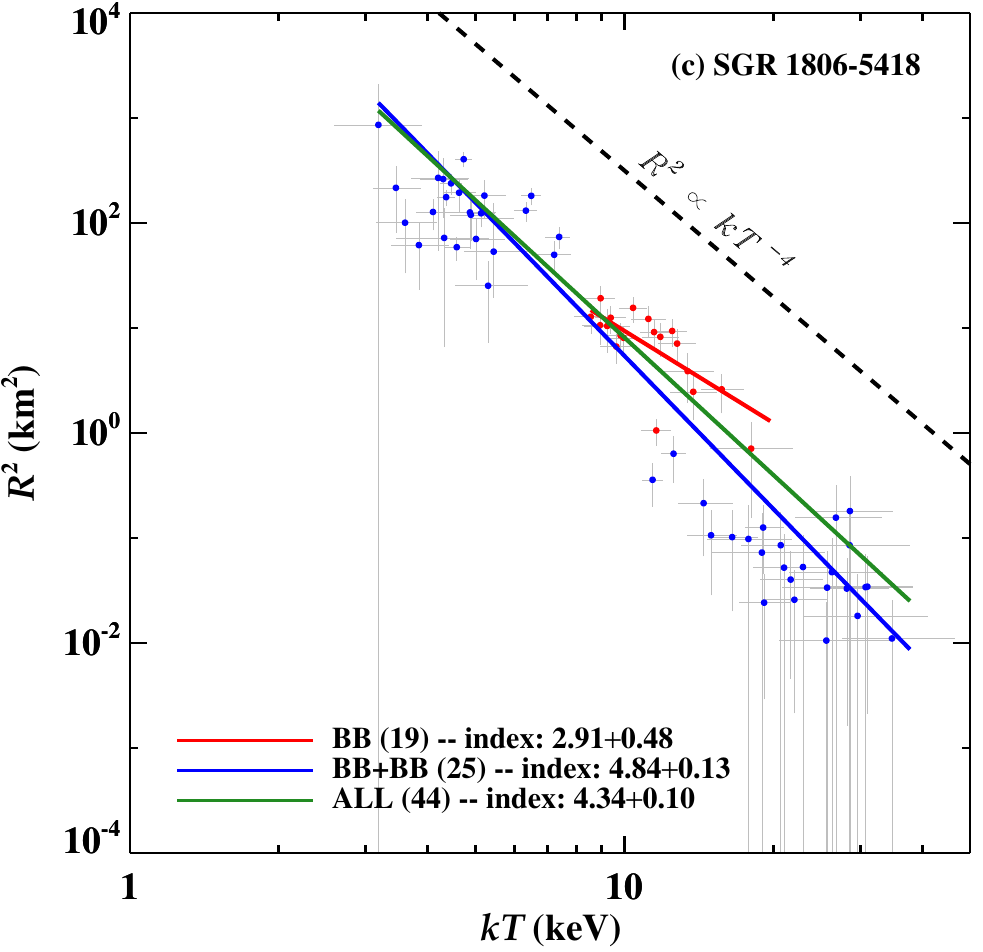}
    \includegraphics[width=0.45\linewidth, trim=60 75 80 260, clip]{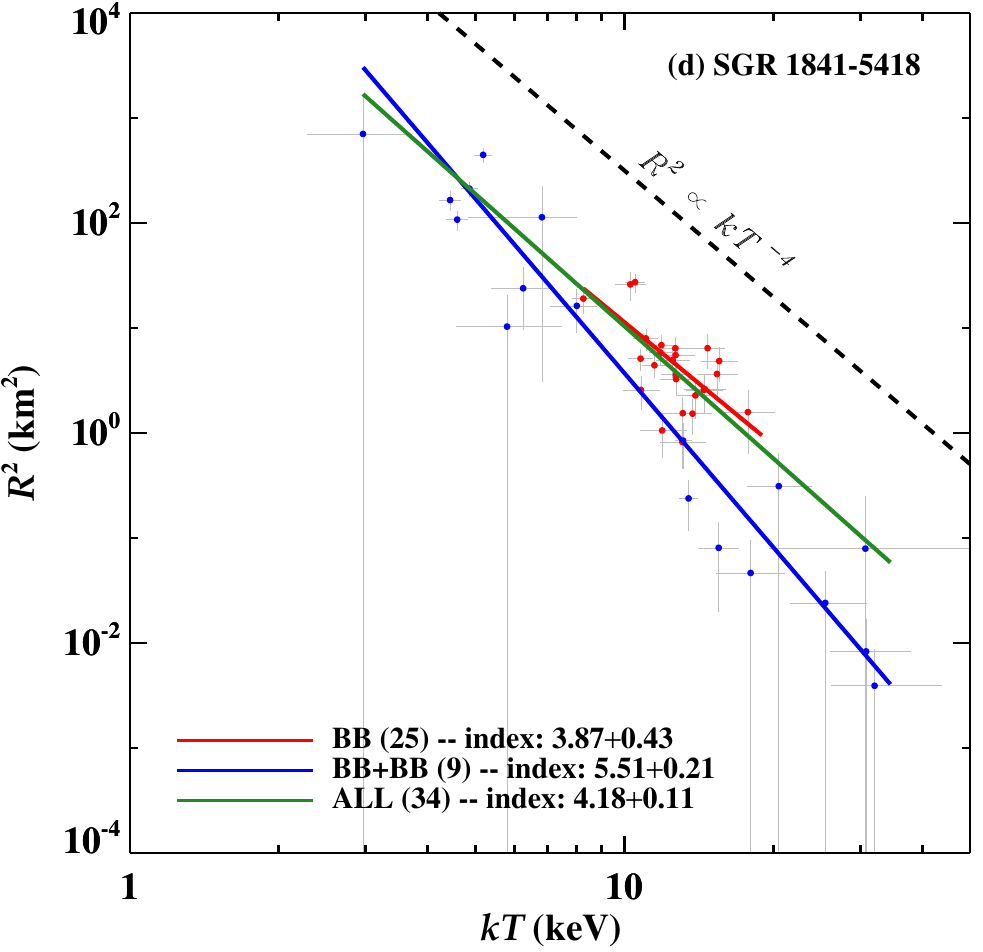} \\
    \caption{$R^2$ vs. $kT$ trends of the burst spectra that only preferred the single BB (red) or BB+BB model (blue) for the sources: (a) \sgronenos, (b) \sgrtwonos, (c) SGR 1806$-$20, and (d) 1E 1841$-$045. The best-fitting power-law model is overlaid on each distribution.}
    \label{fig:r2vskt_bb_2bb}
\end{figure}

\newpage

\section{Summary}

This catalog represents the longest running, high sensitivity, all-sky observations of magnetar events using \fermi data from 2008 to 2025, containing 1254 bursts observed from 17 known Galactic sources. The data is made possible by \ferminosp's exceptional spectral and temporal capabilities of detecting magnetar bursts. We present our methodology for data selection, localization, burst duration estimation, as well as systematic spectral analysis model preference. This catalog extends the scope of the 5-year catalog by \citet{Collazzi2015} by a factor of three in the number of bursts. Moreover, the list of 19 unidentified bursts presented in \citet{Collazzi2015} is now attributed to known sources based on better localization using more advanced tools\footnote{Gamma-ray Targeted Search:  \url{https://github.com/USRA-STI/gamma-ray-targeted-search}} and concurrent \textit{Swift} BAT observations. The data products used for our spectral and temporal investigations are made publicly available\footnote{\url{https://magnetars.sabanciuniv.edu/sgr_catalog/}, the Fermi Science Support Center (FSSC) \url{https://fermi.gsfc.nasa.gov/ssc/data/access/gbm/sgr}, and Zenodo repository:\dataset[doi:10.5281/zenodo.21351728] {https://doi.org/10.5281/zenodo.21351728} \citep{zenodo-catalog}}.

The next generation missions, such as wide-field X-ray monitors, high-time resolution gamma-ray detectors, and advanced X-ray polarimeters, can benefit from our catalog, along with the recently released catalogs from other missions \citep{Chu2026,Pacholski2026}. These catalogs with more complete distributions of magnetar burst properties, would not only reduce uncertainties in the classification and source identifications but also help develop more sophisticated onboard triggering schemes for future instruments that might have the capabilities of polarization, finer time resolution, and coordinated multiwavelength observations.

\begin{acknowledgments}

We thank the reviewer for the thorough report and constructive comments. We also thank the members of Fermi GBM team for maintaining the operations of the instruments that made this research possible. M.G. and the UAH coauthors gratefully acknowledge NASA funding from cooperative agreement 80NSSC25M0084. O.J.R. gratefully acknowledges NASA funding through contract 80MSFC17M0022 and Research Ireland Pathway Funding through contract 24/PATH-S/12742(T). E.G., Y.K., and \"O.K. gratefully acknowledge the support from the Scientific and Technological Research Council of Turkey (T\"UB\.ITAK project number 121F266). D.M.P. gratefully acknowledges NASA funding under Agreement No. 80NSSC23K0552 issued through the Office of Science. M.G.B. thanks NASA for generous support under awards 80NSSC22K0777 and 80NSSC22K1576. In accordance with Federal law, New Mexico Consortium is prohibited from discriminating on the basis of race, color, national origin, sex, age, or disability. E.P. gratefully acknowledges the financial support provided by the National System of Researchers (SNII-SECIHTI) through a research assistant scholarship (SNII III).
\end{acknowledgments}
\newpage
\bibliography{refs}
\bibliographystyle{aasjournal}

\appendix

\section{Effective Exposure Times of the 17 Sources} \label{app_a}

The \emph{Fermi} satellite makes $\sim$15 orbits around the Earth every day. Depending on its orbital path, it passes through the South Atlantic Anomaly (SAA), during which all detectors are turned off to avoid damage from excessive charged particle flux, and sources enter and exit Earth occultation. To calculate the effective GBM exposure times for each of the 17 magnetars in our catalog, we utilized \fermi position history files, together with the coordinates of each source. For each day within our catalog span (2008 $-$ 2025), we calculated the cumulative time the source remained above the Earth's limb, while excluding the time ranges of the spacecraft's passages through the SAA. Table \ref{tab:avg-exposure} presents the calculated average daily effective exposure for each magnetar included in this catalog.

\begin{table}[!htbp]
    \centering
    \caption{The average daily exposure times per source}
    \label{tab:avg-exposure}
    \begin{tabular}{lc}
    \hline\hline
    \colhead{\bf Source} & \colhead{\bf Average exposure time (s)} \\
\hline
        4U 0142+61& 60605 \\
        1E 2259+586& 59740 \\
        SGR 0418+5729& 59331 \\
        SGR 0501+4516& 54786 \\
        1E 1048.1-5937& 55480 \\
        SGR J1550-5418& 54053 \\
        SGR J1555.2-5402& 53986 \\
        CXOU J164710.2-455216& 51671 \\
        1RXS J170849.0-400910& 49437 \\
        SGR J1745-2900& 47526 \\
        SGR 1806-20& 46849 \\
        Swift J1818.0-1607& 46642 \\
        Swift J1822.3-1606& 46647 \\
        SGR 1830-0645& 46475 \\
        1E 1841-045& 46478 \\
        PSR J1846-0258& 46495 \\
        SGR J1935+2154& 47942 \\ \hline
    \end{tabular}

\end{table}

\end{document}